\documentclass[a4paper,fleqn,usenatbib]{mnras}
\usepackage{newtxtext,newtxmath}

\usepackage[T1]{fontenc}
\usepackage{ae,aecompl}

\usepackage{graphicx}	
\usepackage{amsmath}	
\usepackage{amssymb}	
\usepackage{times}
\usepackage{gensymb}
\usepackage{pifont}
\usepackage{placeins}
\usepackage{textcomp}

\usepackage{longtable}

\title[The kinematics of OB associations in {\it Gaia}-DR2]{Not all stars form in clusters -- {\it Gaia}-DR2 uncovers the origin of OB associations}

\author[J. L. Ward, J. M. D. Kruijssen \& H.-W. Rix]{
Jacob L. Ward\thanks{E-mail: jakelward1@gmail.com (JLW)}$^{1}$, J. M. Diederik Kruijssen$^{1}$ and Hans-Walter Rix$^{2}$
\\
$^{1}$Astronomisches Rechen-Institut, Zentrum f\"{u}r Astronomie der Universit\"{a}t Heidelberg, M\"{o}nchhofstra{\ss}e 12-14, D-69120 Heidelberg, Germany \\
$^{2}$ Max Planck Institute for Astronomy, K\"{o}nigstuhl 17, D-69117 Heidelberg, Germany
}

\date{Accepted XXX. Received YYY; in original form ZZZ.}

\pubyear{2020}

\begin{document}
\label{firstpage}
\pagerange{\pageref{firstpage}--\pageref{lastpage}}
\maketitle

\begin{abstract}
Historically, it has often been asserted that most stars form in compact clusters. In this scenario, present-day gravitationally-unbound OB associations are the result of the expansion of initially gravitationally-bound star clusters. However, this paradigm is inconsistent with recent results, both theoretical and observational, that instead favour a hierarchical picture of star formation in which stars are formed across a continuous distribution of gas densitiesand most OB associations never were bound clusters. Instead they are formed in-situ as the low-density side of this distribution, rather than as the remnants of expanding clusters. We utilise the second {\it Gaia} data release to quantify the degree to which OB associations are undergoing expansion and, therefore, whether OB associations are the product of expanding clusters, or whether they were born in-situ, as the large-scale, globally-unbound associations that we see today. We find that the observed kinematic properties of associations are consistent with highly substructured velocity fields and additionally require some degree of localised expansion from sub-clusters within the association. While most present-day OB associations do exhibit low levels of expansion, there is no significant correlation between radial velocity and radius. Therefore, the large-scale structure of associations is not set by the expansion of clusters, rather it is a relic of the molecular gas cloud from which the association was formed. This finding is inconsistent with a monolithic model of association formation and instead favours a hierarchical model, in which OB associations form in-situ, following the fractal structure of the gas from which they form.
\end{abstract}




\section{Introduction}

It is widely accepted that most stars (70--90\%) form in groups of at least 35 stars with densities $>$ 1\,$M_{\sun}$pc$^{-3}$ \citep{LadaLada2003}. Historically, this result has often been used to assert that most (if not all) stars form in gravitationally-bound clusters. In this monolithic formation scenario, the currently graviationally-unbound nearby OB associations (e.g. \citealt{Melnik2017}) must have been significantly more compact, either as a single cluster (singularly monolithic) or multiple clusters (multiply monolithic), at the time of their formation and must have subsequently expanded into the configurations we see today (e.g. \citealt{Lada1991,Brown1997,Kroupa2001}). The most commonly invoked mechanism for this process is the expulsion of residual gas from embedded clusters through stellar feedback, rendering the clusters super-virial (e.g. \citealt{Hills1980,Goodwin2006,Baumgardt2007}). This monolithic view of star formation seems somewhat contradictory to observations of present day star formation proceeding in a wide range of environments including large-scale hierarchical structures and isolated YSOs (e.g. \citealt{Gomez1993,Allen2007,Gutermuth2008,Evans2009,Lamb2010}).

The effectiveness of gas-expulsion as a mechanism for cluster disruption has been called into question by recent numerical work \citep{Kruijssen2012c, Girichidis2012,Dale2015}. These studies are supported by recent observations that suggest that proto-clusters in W51 are evolving towards a state of gas-exhaustion rather than gas-expulsion \citep{Ginsburg2016}. Moreover, observations of young massive cluster (YMC) progenitors have concluded that the cluster mass must become more centrally concentrated than the gas reservoirs from which they form (\citealt{Walker2015,Walker2016}, see also the review of \citealt{Longmore2014}), again suggesting that gas-exhaustion has a greater impact on the dynamical evolution of young clusters than gas-expulsion.  Detailed models of the early evolution of embedded clusters by \citet{Sills2018} find that gas does not have a significant effect on the motions of stars in the earliest stages of cluster evolution. However, these models do not include stellar feedback, so they are unable to assess the effectiveness of feedback-driven gas expulsion. 

In \citet{Paper1}, we used the Tycho-{\it Gaia} Astrometric Solution (TGAS) to determine four kinematic properties of 18 nearby OB associations in order to measure the extent to which they are currently expanding and therefore judge whether or not they could have been produced by the monolithic model of star formation outlined above. We found that none of the associations exhibited any significant evidence of an expanding velocity field and that the bulk kinematic properties of the sample are far more consistent with randomised velocity fields than expanding ones. Importantly, we also found that low levels of apparent expansion, in the form of radial velocities of a few km\,s$^{-1}$ can easily be reproduced by a randomised velocity field where positional sub-structure is present.

Recent analyses of the dynamics of OB associations have yielded similar results. The detailed investigation into the origin and evolution of the OB association Cygnus OB2 by \citet{Wright2014,Wright2015,Wright2016} found no evidence of a violent relaxation event and instead found that Cygnus OB2 was likely always an association and has not undergone significant dynamical evolution. Similarly, \citet{Wright2018} used the first {\it Gaia} data release to show that the Sco-Cen OB association was most likely formed in a highly substructured state with multiple small-scale star formation events rather than a single, monolithic burst of star formation. Another study of 18 nearby OB associations (although different associations to those analysed in \citealt{Paper1}), using the first {\it Gaia} data release found little evidence of systematic expansion of OB associations \citep{Melnik2017} and a subsequent study of 28 associations found that the majority of associations are not undergoing expansion \citep{Melnik2020}. \citet{Cantat-Gaudin2018} find that, while the Vela OB2 association is expanding, the expanding stellar distribution suggests that this expansion began before the stars in Vela OB2 were formed, probably the result of a supernova. Vela OB2 is therefore a clear example of an OB association that was formed globally unbound but still exhibits signs of expansion. For a review of unbound young stellar systems see \citet{Gouliermis2018}.

In this second paper, we extend our original study to the second {\it Gaia} data release ({\it Gaia}-DR2). We apply a clustering algorithm in order to self-consistently identify nearby (distance $< 4$\,kpc) OB associations and their members. We then quantify the same kinematic properties as investigated in \citet{Paper1} (see Section \ref{results_section}), along with an additional, powerful diagnostic: the relation between radial velocity and radius.
In the next Section, we outline the catalogues used to carry out this study and the application of a clustering algorithm to identify OB associations and their members. We then quantify the kinematic properties of the selected OB associations in Section \ref{results_section} and in Section \ref{improved_models} we introduce new models in order to reproduce the observed properties of OB associations. We then discuss the implications of this work, alongside a discussion of the caveats of our work, in Section \ref{discussion}. Finally, we summarise our results and the outlook for future studies in Section \ref{conclusions}.

\section{Sample selection, data reduction and model distributions}

The second {\it Gaia} data release ({\it Gaia}-DR2) represents the first five-parameter (positions in the plane of the sky, parallaxes, and proper motions) data release produced using data exclusively from the {\it Gaia} mission. As such, it is the largest and most precise all-sky astrometric survey to date. While the Tycho-{\it Gaia} Astrometric Solution (TGAS, \citealt{GAIADR12016,Michalik2015}) contains around 2 million stars with five-parameter solutions, {\it Gaia}-DR2 contains over 1.3 billion \citep{Gaia2018a,Gaia2018b}.

We cross-reference the {\it Gaia}-DR2 catalogue with the Galactic OB star catalogue (GALOBSTARS; \citealt{Reed2003,Reed2005}). This is the largest all-sky catalogue of OB type stars, adding $\sim$5500 objects to the Case-Hamburg Galactic Plane Luminous Stars surveys \citep{Hardorp1959,Hardorp1964,Hardorp1965,Nassau1963,Nassau1965,Stock1960,Stephenson1971}, yielding a total of over 16000 known and high-probability OB-star candidates \citep{Reed2003}. Main sequence stars are selected as OB-type stars if they exhibit a spectrum or $UBV$-based photometrically determined $Q-$value consistent with that of spectral type B2V or earlier. For more luminous (I-IV) classifications, spectral types are accepted as late as B9 and the $Q-$value limit imposed is consistent with supergiants of type B6. 

\subsection{Selecting OB stars in \it{Gaia}-DR2}

\begin{figure}
	\includegraphics[width=1.0\linewidth]{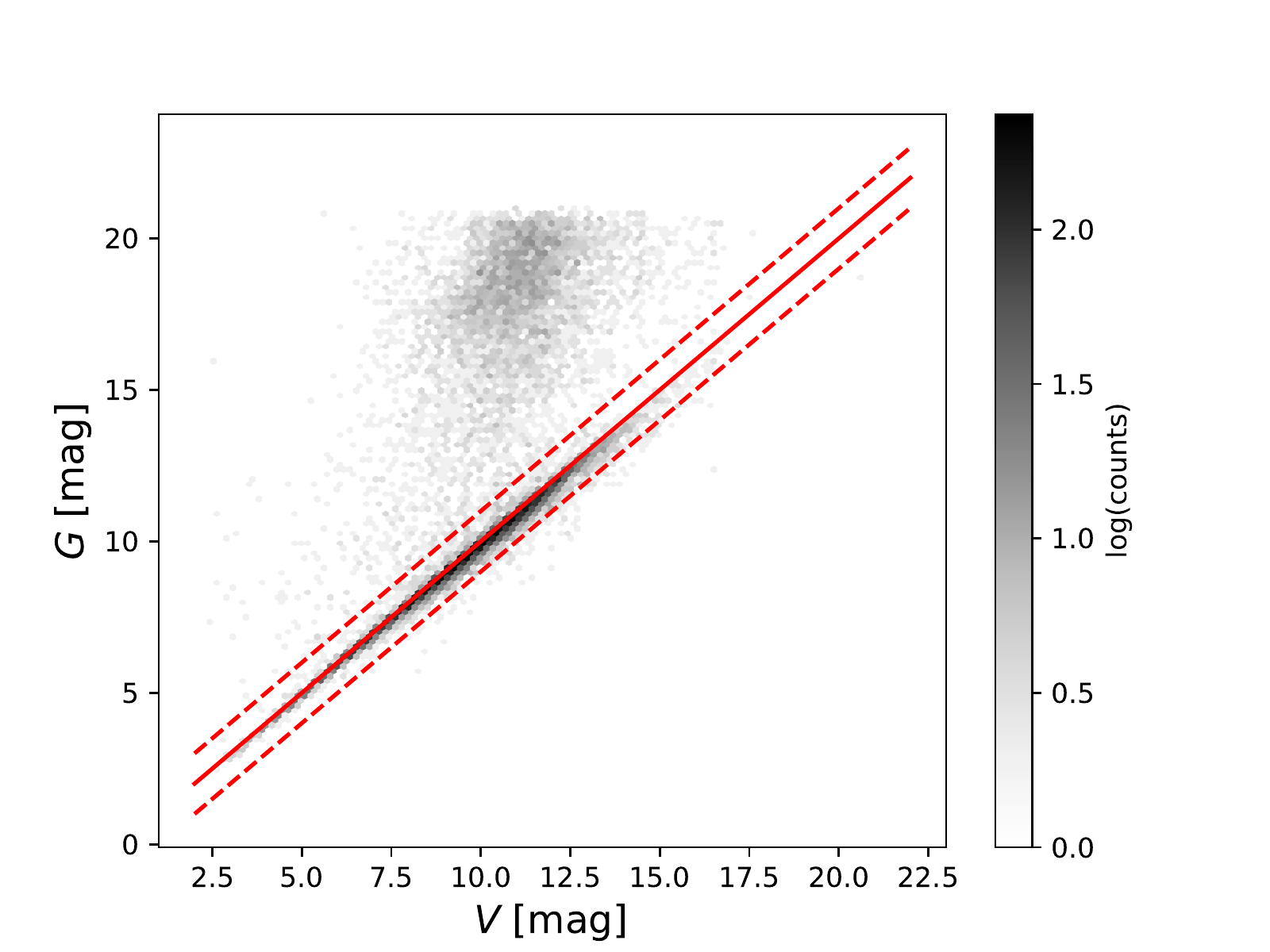}
	\caption{\label{VGfig} $G-$band magnitude versus $V-$band magnitude for all of the sources matched in {\it Gaia-}DR2 with sources in the GALOBSTARS catalogue. The solid red line is a one-to-one line and the two dashed lines mark the region from which we select our sample. Note the strong concentration of genuine OB stars around the one-to-one relation, as expected from the approximately zero $V-G$ colour of OB stars. Those sources that exhibit significantly higher $G-$band magnitudes are most likely to be lower mass contaminants that represent a mismatch between the catalogues.}
\end{figure}

\begin{figure*}

		\begin{center}
		\includegraphics[width=0.49\linewidth]{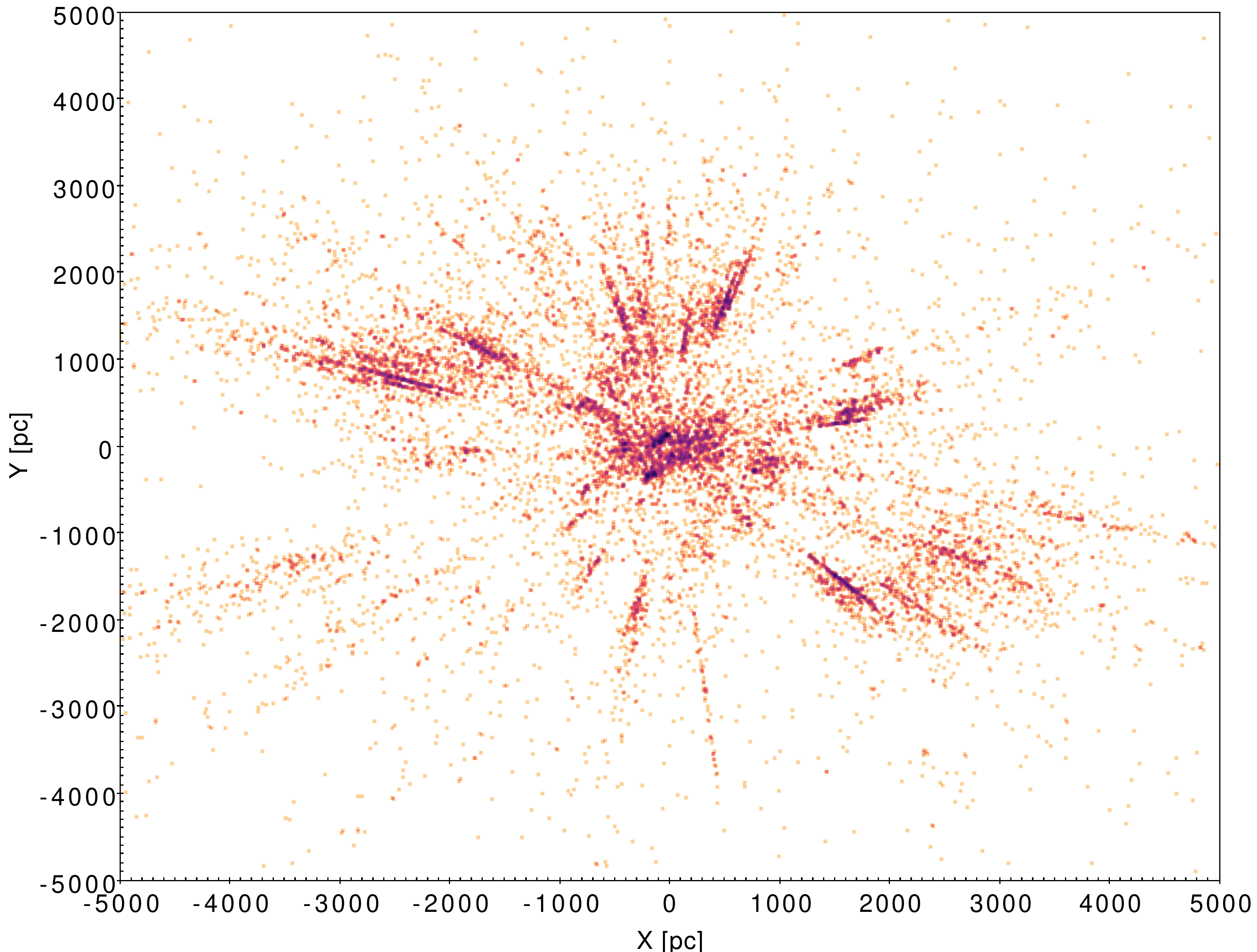}
		\includegraphics[width=0.49\linewidth]{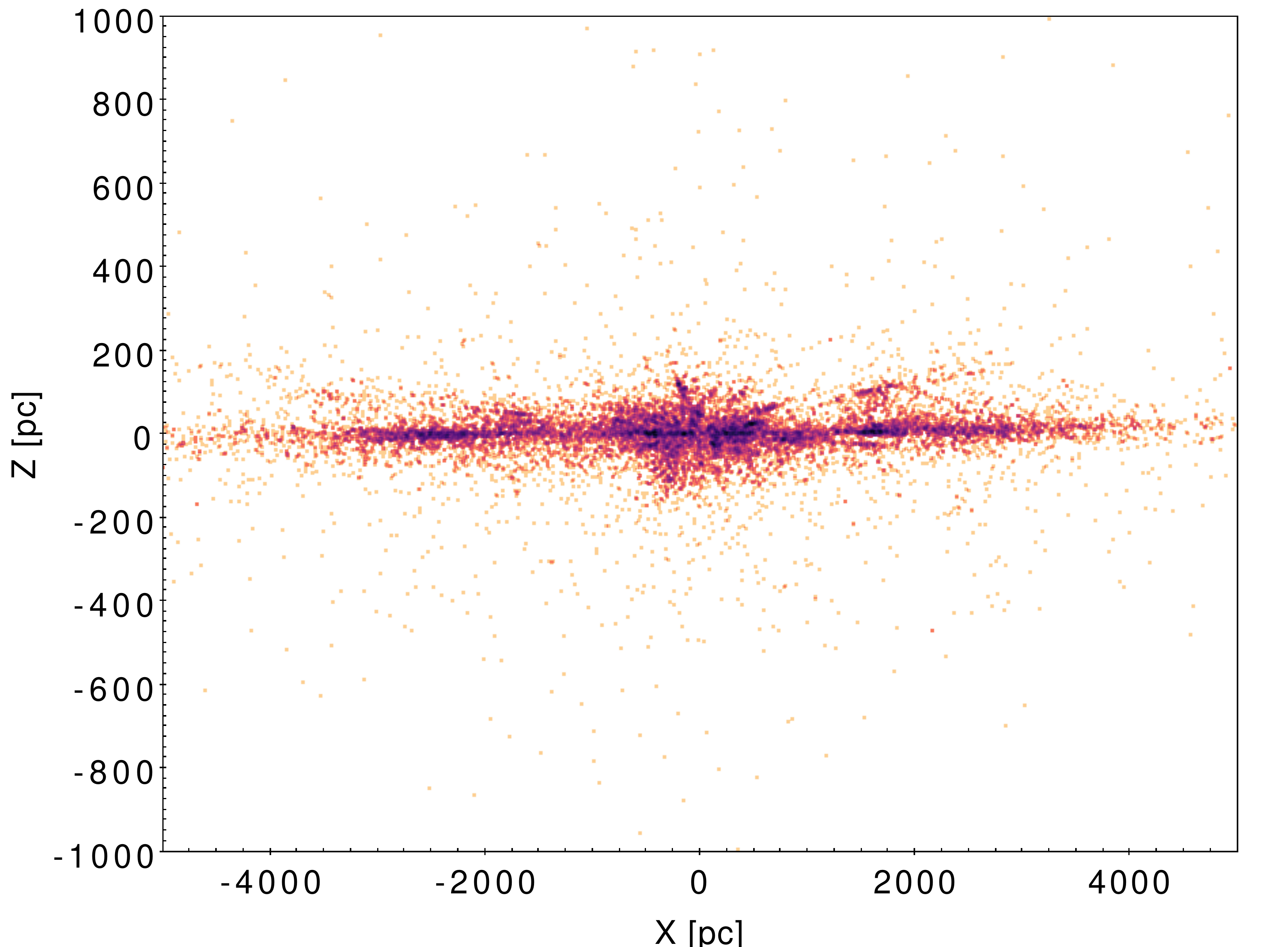}
		\end{center}
	\caption{\label{spokey_fig} Distribution of cross-matched OB-type stars relative to the Sun. Note that the apparent elongation of the associations in the radial direction is primarily due to the uncertainty in parallax towards those stars. Here the definition of the axes X,Y, and Z axes is an arbitrary choice without bearing on the subsequent analysis}

\end{figure*}

First we cross-match\footnote{Cross-matching is carried out with the {\sc Topcat} \citep{Taylor2005} Table Access Protocol (TAP) interface.} the GALOBSTARS catalogue with the {\it Gaia}-DR2 catalogue ( translated into the J2000 epoch) in RA and Dec by selecting the all matches within a 5\,arcsec radius of the coordinates reported in the GALOBSTARS catalogue. This relatively wide 5\,arcsec radius is used because the GALOBSTARS catalogue is a composite of a number of source catalogues of varying precision and may contain a variety of unaccounted for systematic effects. We then compare the {\it V}-band magnitudes from the GALOBSTARS catalogue with the {\it G}-band magnitudes from the {\it Gaia}-DR2 catalogue as shown in the left panel of Fig. \ref{VGfig}. As expected, there is a strong correlation between $V-$band and $G-$band magnitudes where $G\approx V$. We therefore apply a cut in $V-G$ space, excluding any stars that do not meet the criterion $-1<V-G<1$. Through this cut we have excluded  the majority of lower mass contaminants with $V-G<1$, and those massive stars that are the most embedded with $V-G>1$. This helps to ensure that the stars selected as OB stars in the {\it Gaia}-DR2 catalogue are indeed the same stars that appear in the GALOBSTARS catalogue. Removal of the most embedded sources is carried out in order to study the kinematics of stellar structures post gas-expulsion. After applying these cuts, 11844 stars remain in the sample. The distribution of the selected OB stars with respect to the Sun is shown in Fig. \ref{spokey_fig}. This is shown in the Cartesian coordinate space used in the following section to select probable OB associations.

In order to test the likelihood of chance projection effects of lower-mass stars with similar apparent magnitudes as the objects in the GALOBSTARS catalogue contaminating our sample, we carry out the same cross-matching process using the mock {\it Gaia--}DR2 catalogue of \citet{Rybizki2018}. For the 18693 stars in the GALOBSTARS catalogue with O- and B-type classifications, we obtain matches within 5\,arcsec for 18959 stars in the mock {\it Gaia--}DR2 catalogue. Performing the same $V-G$ magnitude cut as above for the sample of 13793 matches where $V-$band magnitudes are present in the GALOBSTARS catalogue, results in a total of 25 matches that would satisfy the selection criteria in this work. This corresponds to a contamination rate of $<0.2$ per cent. As low-mass stellar contaminants must lie at shorter distances in order to exhibit the same apparent magnitude as an OB-type star, we can expect that very few, if any, low-mass stars contaminate the samples of OB stars that are assigned to associations in the following section.

Distances are obtained for each selected OB-type star from the catalogue of distances of \citet{Bailer-Jones2018}. This catalogue contains distances towards 1.33 billion stars inferred from the parallaxes published in {\it Gaia}-DR2 using a weak distance prior that varies smoothly as a function of Galactic longitude and latitude. While more precise distance estimates are possible using auxiliary data, this catalogue provides the best distances available independent of assumptions of interstellar extinction and stellar properties, and is therefore the most appropriate determination of distances to apply to an all-sky study. The comparison of \citet{Anders2019} between distances derived by \citet{Bailer-Jones2018} and more precise astro-spectrometric distances determined using the {\sc StarHorse} code \citep{Santiago2016,Queiroz2018} shows that for the majority ($>$90\%) of stars up to a distance of 3\,kpc, there is no systematic difference in the derived distance. The distances derived by \citet{Bailer-Jones2018} tend to be shorter due to the exponentially decreasing prior that tends to constrain most stars to within 6\,kpc. \citet{Anders2019} conclude that their results are not more precise than those of \citet{Bailer-Jones2018} but are more accurate at large distances ($>$3\,kpc) due to the more informed prior used. As only 13 of our selected associations are more distant than 3\,kpc, we consider that the distances derived by \citet{Bailer-Jones2018} are sufficiently accurate for the current study. While extinction does not factor into the distance determination for individual stars in \citet{Bailer-Jones2018}, it is included in the Galactic model used to inform the prior as described in \citet{Rybizki2018}. Given that the distance has only an effect on one of the five kinematic diagnostics using in this study (see Section \ref{results_section}), homogeneous treatment of the sample is preferable to obtaining the most precise distances for individual stars.

\subsection{Identifying OB associations in DR2}

\label{assoc_select}

Following the selection of OB stars within the {\it Gaia}-DR2 catalogue, we apply a clustering algorithm to assist in selecting likely OB associations, identified as over-densities in 5-dimensional position-velocity space.

In order to robustly apply a clustering algorithm, the dimensions over which the algorithm is run must be equivalent. Taking the example of standard astrometric coordinate systems of RA, Dec, and parallax, blindly applying a clustering algorithm will always produce a biased result due to the anisotropic nature of parallaxes and positions in the plane of the sky. In order to carry out an unbiased selection, an isotropic metric is required. The obvious choice is therefore to transform the data from RA, Dec and parallax into a 3-dimensional Cartesian coordinate space in which the unit vectors in each of the three coordinates are equivalent. A similar non-equivalence is present between position and velocity spaces, ruling out an unbiased, simultaneous application of a clustering algorithm on a mixed position-velocity coordinate system. 
We therefore opt to remap our sample of OB stars into a Cartesian coordinate system from the spherical coordinate system of RA, Dec and distance. As the orientation of the Cartesian space is an arbitrary decision, we place the OB stars into the Galactic coordinate system as shown in Fig\,\ref{spokey_fig}.

\begin{figure*}
\begin{minipage}{170mm}
	\begin{center} 
	\includegraphics[width=0.9\linewidth]{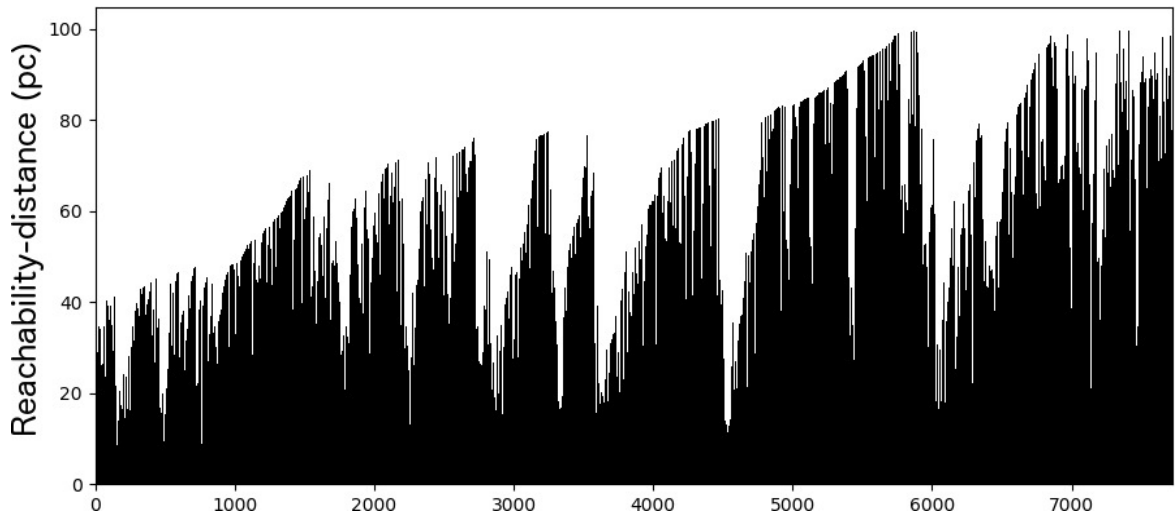}
	\end{center}
	\caption{\label{OPTICS_out} Reachability plot for the initial implementation of the OPTICS clustering algorithm using a maximum reachability distance of 100\,pc and a minimum number of points of 5. The reachability distance is effectively the smallest distance such that a star can be reached by a core object. An object can be considered a core object if a minimum number of stars lie within a specified distance from that object. The stars are arranged such that each point neighbours its closest stars spatially. The clustered structures are represented by troughs in this reachability diagram.}
\end{minipage}
\end{figure*}
By transforming the positions on the sky of a large population of OB stars into a Cartesian coordinate system, we are able to employ a clustering algorithm to identify OB associations using the state-of-the-art astrometric data that {\it Gaia} provides. In this way, our study is insensitive to an over-reliance on previously obtained association catalogues and traditional definitions of OB associations. 
We can then independently select over-densities of OB stars in an automated way. The selection function is therefore only dependent on the two catalogues used ({\it Gaia}-DR2 and GALOBSTARS), and the cluster identification algorithm implemented. The initial identification of over-densities of OB stars does not take velocity structure into account. We make the assumption that, due to the inherent youth of massive stars, OB stars that are clustered in position space were likely formed in the same environment. However, the velocity structure is taken into account when identifying members of each association later in this section.

To identify over-densities of OB associations, we use the OPTICS (Ordering Points To Identify the Clustering Structure) algorithm \citep{Ankerst1999}. OPTICS is a density-based cluster-ordering algorithm, producing an augmented ordering of the database representing its density-based structure.  OPTICS re-orders a database, storing a \textquoteleft reachability-distance\textquoteright\, for each entry in the database. The reachability distance is defined as the shortest distance such that the point is directly density-reachable from another point if that second point neighbours at least some minimum number of neighbours ($N_{\text{min}}$) within some maximum distance ($\epsilon$). The reachability of some point $p$ with respect to another point $o$ is defined as:
\begin{equation}
d_{\text{reach}}(p,o) =
\begin{cases}
\text{Undefined},& \text{if } |N_{\epsilon}(o)| < N_{\text{min}} \\ 
\text{max}[d_{\text{core}}(o),d(o,p)],& \text{otherwise}
\end{cases}
\end{equation}
where $N_{\epsilon}(o)$ is the neighbourhood of $o$ containing all points within the maximum distance ($\epsilon$) of point $o$. The core distance, $d_{\text{core}}$ for a point $o$ is defined as:
\begin{equation}
d_{\text{core}}(o) = 
\begin{cases}
\text{Undefined},& \text{if } |N_{\epsilon}(o)| < N_{\text{min}} \\ 
d_{\text{minpt}}(o),& \text{otherwise}
\end{cases}
\end{equation}
where $d_{\text{minpt}}(o)$ is the smallest distance $\epsilon^{\prime}$ between $o$ and a neighbouring point such that $o$ would be a core object with respect to $\epsilon^{\prime}$ if the neighbour lies within the neighbourhood $N_{\epsilon}(o)$. A point is defined as a core object if it meets the condition that the number of points within the distance $\epsilon$ is greater than the minimum number of points required, i.e. $|N_{\epsilon}(o)| > N_{\text{min}}$. 

Rather than directly identifying clusters, OPTICS creates an augmented ordering of clustered data representing its density-based clustering structure, in which points are ordered such that each point falls between its closest neighbours in the coordinate system to which the algorithm is applied. Clustered structures can then be extracted from the resulting ordered database, represented by troughs in the reachability diagram shown in Fig. \ref{OPTICS_out}. Clusters appear as troughs in the reachability diagram because an overdensity in positional space implies a shorter reachability distance for stars within the overdensity. The key advantage of OPTICS over other clustering algorithms such as DBSCAN is its insensitivity to varying levels of noise and background densities, making OPTICS uniquely well-suited to all-sky applications amongst clustering algorithms. OPTICS assumes nothing about the shape of a particular overdensity, and is therefore well suited to the identification of structures that are gravitationally-unbound and therefore are not necessarily spherically symmetric. The OPTICS algorithm has rarely been previously applied to astronomical problems although where it has been applied it has proved to be a useful means of identifying clustered structure (e.g. \citealt{SansFuentes2017,Canovas2019}).

We apply the OPTICS clustering algorithm in an iterative fashion, selecting over-densities with a minimum of five stars, in order to take into account any hierarchical structure with the aim of identifying the smallest scale over-densities. This is so that we identify the most cluster-like groups that are likely to exhibit expanding velocity fields if formed through the monolithic model as described in Section 1. This iterative application of OPTICS is carried out by initially running OPTICS with a maximum reach length of $\epsilon = $100\,pc and decreasing $\epsilon$ in subsequent iterations. The results of the initial OPTICS application are visualised in the reachability plot shown in Fig. \ref{OPTICS_out}. This reachability plot shows the reachability distance for every star, ordered by the OPTICS algorithm. Over-densities in stellar populations are represented by the troughs in the reachability diagram. Those over-densities consisting of fewer than 10 stars are rejected.

These initially selected over-densities are subsequently used as input catalogues for the next iterations of the OPTICS algorithm, reducing the maximum reach length to $\epsilon =$\,90\,pc. If no new over-densities with at least 10 stars are found, then the parent level overdensity is selected as an association. Otherwise, the new over-densities are added to the list of OPTICS input files.
This process is then repeated, reducing the maximum reach length by 10\,pc for each cycle untill a final run using a maximum reach length of $\epsilon = $10\,pc. We use a smallest maximum reach length of $\epsilon = $10\,pc because reducing this any further increases the risk of arbitrarily separating associations into substructures, while imposing a maximum length of 10\,pc does not prevent the identification of structures that are smaller than this scale. This process has resulted in the identification of 110 potential OB associations.

Having identified 110 potential OB associations in Cartesian space, we must identify additional, lower mass members of the associations. In order to do this, we obtain data for all stars within the vicinity of the associations in RA, Dec, parallax space from the {\it Gaia}-DR2 archive. One association (assoc. 90, Sco-Cen) is removed from the sample as this association extends over a large region of the sky and is known to host multiple, kinematically distinct populations \citep{Wright2018}, and cannot necessarily be considered as a single association. We then determine a maximum distance between one OB star and its nearest neighbouring OB-type star in the plane of the sky and remove any stars that do not fall within this distance from the sample. We then obtain the most likely distances for the remaining sample and use those to place all stars into a 5-dimensional Cartesian position-velocity coordinate space. We remove the most extreme outliers ($v > 3\sigma_{v}$ in either velocity dimension) from the sample of OB-type stars in velocity space. Then we calculate the distance for each OB star and its nearest neighbouring OB star. We exclude any stars that do not fall within that distance for their nearest OB-type neighbour. This process is performed in both 3-dimensional position space and 2-dimensional velocity space.

\subsection{Properties of observed OB associations}

\begin{figure*}
	\includegraphics[width=\linewidth]{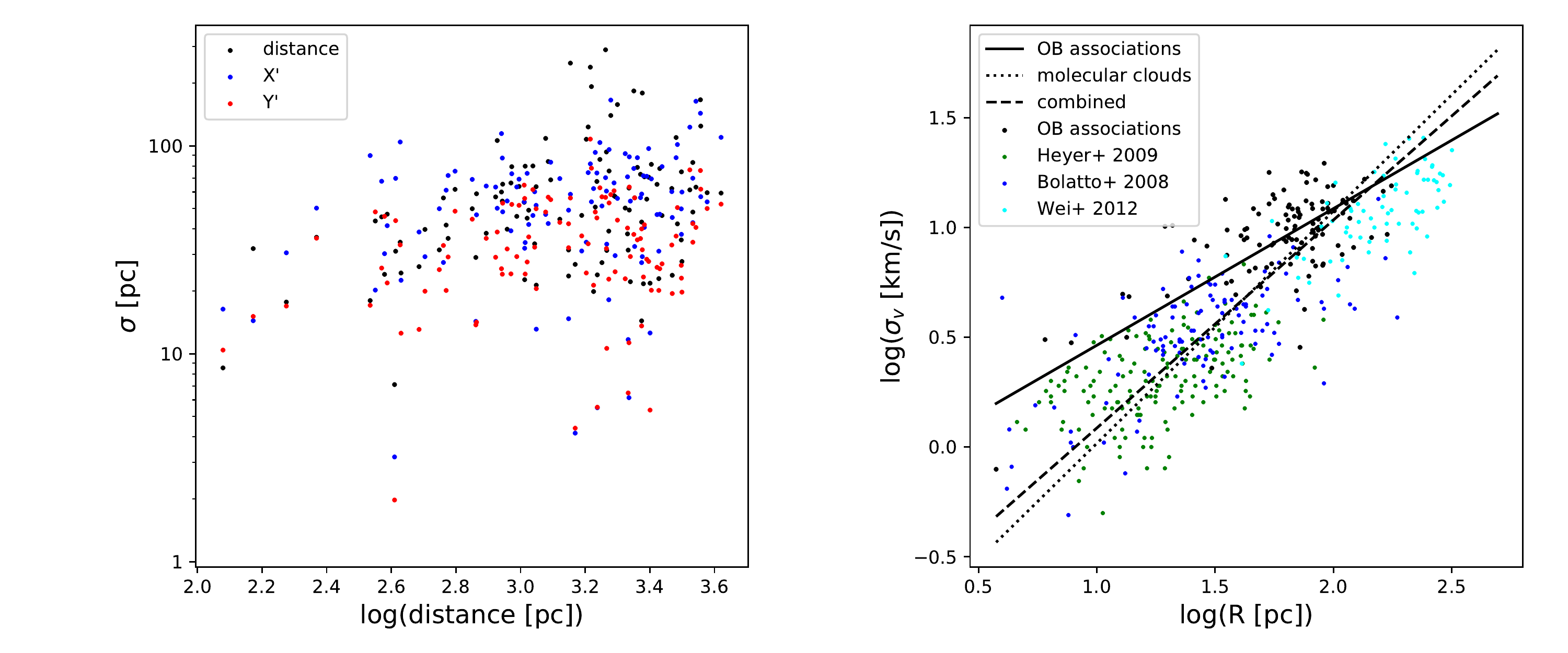}
	\caption{\label{size-linewidth_fig} Left: The standard deviation of the line-of-sight coordinate in each association as a function of distance along the line of sight (black) as well as that of the two orthogonal axes in the plane of the sky aligned with right ascension (X$^{\prime}$, blue) and declination (Y$^{\prime}$, red) at the mean RA and Dec of the association. The median size of the associations is $\sigma \approx 50$\,pc. Right: The two-dimensional velocity dispersion in the plane of the sky normalised by $\sqrt{2}$ versus the radius for all OB associations. Also shown are the line-of-sight velocity dispersion versus radius data for Galactic molecular clouds analysed by \citet{Heyer2009} in green, molecular clouds in nearby galaxies from \citet{Bolatto2008} in blue, and molecular clouds in the Antennae galaxies from \citet{Wei2012} in cyan. The linear Thiel-Sen \citep{Sen1968} regression fit to the OB associations is shown in the solid black line with the fits to the molecular clouds and the combined data set of OB associations and molecular clouds shown in the dotted and dashed lines, respectively.}
\end{figure*}

We measure the spatial dispersions in the plane of the sky and along the line of sight, as well as the velocity dispersions in the plane of the sky for each of the associations in our sample.
These measurements are presented in Table \ref{app_tab1}. The left-hand panel of Fig. \ref{size-linewidth_fig} shows the three dispersions in position space as a function of distance. While there is a slight tendency for the associations at greater distances to exhibit higher dispersions, the correlation is extremely weak. This is an indication that the dispersion measurements in positional space presented in Table \ref{app_tab1} are not dominated by the uncertainty in parallax, as such an effect would manifest in higher dispersions at greater distances (smaller parallaxes). Note, that towards some of the more distant associations, the dispersions in distance become significantly larger than in the plane of the sky, suggesting that a number of associations may be unresolved along the line of sight. However, the majority of the kinematic diagnostics used in this work (See Section 4) are independent of distance so this does not represent a significant concern.

The selected associations span a range of mean distances from less than 200\,pc to over 4\,kpc. Note that these distances are the mean of distance determinations for individual stars and should not be treated as accuarate distances for each association. The lowest number of stars selected is 40 stars (association 79) with the highest $\sim$1.8$\times$10$^{5}$ stars (association 48) with mean and median sample sizes of $\sim$16000 and $\sim$6000, respectively. The proportion of association members identified as OB-type stars range from 0.01 per cent (association 12) to 100 per cent (association 79), with a mean of 2 per cent and a median of 0.3 per cent. None of these association properties are correlated with distance or the position in the plane of the sky, indicating that the sample is not dominated by selection effects.

In the right-hand panel of Fig. \ref{size-linewidth_fig}, we show the two-dimensional velocity dispersion in the plane of the sky normalised by $\sqrt{2}$ versus the radius (defined as $R=\sqrt{\sigma_{X}^{2}+\sigma_{Y}^{2}}$) of each association for the entire sample. Alongside these data, are the line-of-sight velocity dispersion versus radius for molecular clouds in the Milky Way \citep{Heyer2009}, nearby dwarf galaxies \citep{Bolatto2008}, and the Antennae galaxies \citet{Wei2012}. The scatter in the data is large and a number of observed stellar structures fall at higher velocity dispersions than the observed sample of molecular clouds. However, for the most part they follow a similar relation.

We fit a Theil-Sen estimator regression to the data as this is considerably less sensitive to the presence of outliers when compared to least-squares-based fitting methods \citep{Sen1968,Rousseeuw1987}.
Fitting a Thiel-Sen regression to the velocity dispersion versus the radius of each association, we determine the following relations when fitting to the OB associations only, the molecular clouds only, and the combined data set of OB associations and molecular clouds:
\begin{equation}
\log(\sigma_{v})_{\text{OB associations}} = -0.07+0.62\substack{+0.13\\-0.12}\log(R)
\end{equation}
\begin{equation}
\log(\sigma_{v})_{\text{molecular clouds}} = -1.04+1.06\substack{+0.04\\-0.04}\log(R)
\end{equation}
\begin{equation}
\log(\sigma_{v})_{\text{combined}} = -0.69+0.85\substack{+0.03\\-0.03}\log(R)
\end{equation}
The slope fitted to the OB associations without the inclusion of molecular clouds differs significantly from the molecular cloud fit. However, the scatter in both samples is large and for the most part, the OB associations fall within the scatter of the molecular cloud sample. The difference in slope is largely due to a small number of associations with low radii but high velocity dispersions. In comparison, the addition of the associations to the molecular cloud data has a somewhat minor effect on the overall gradient of the relation.
While the scatter is significant, the slope of the fit to the OB associations is similar to the slope of the size-linewidth relation of \citet{Larson1981}  of 0.38 and the slope of 0.39 determined for the substructures of Cyg OB2 and Car OB1 by \citet{Lim2019}. \mbox{\citet{Lim2019}} suggest that the closeness of OB associations to the size-linewidth relation derived for molecular clouds is an indication that star formation across gravitationally-unbound turbulent clouds may be a better explanation for the formation of OB associations rather than the expansion of clusters. However, given the large degree of scatter in Figure \ref{size-linewidth_fig}, we are unable to draw any substantial conclusions regarding the origin of OB associations from this relation alone.

\subsection{Model distributions}

In order to quantify the degree to which the observed associations are consistent with expanding velocity distributions, we produce six sets of model associations representing random velocity fields, and expanding profiles from both a single point and multiple points. These model associations are generated based on the properties of the observed associations, but we change elements of their position-velocity structure to enable an interpretation of the properties of the observed associations. The production of these model associations is described in detail in \citet{Paper1} and the six model cases are briefly summarised here:
\begin{enumerate}
	\item Observed stellar positions and observed absolute velocities with random directions of motion (case I).
	\item Observed stellar positions and observed absolute velocities where $1/3$ of the stars are guaranteed to have positive radial velocities with respect to the mean position and velocity of the $N$ nearest neighbours, introducing a net expansion (case II).
	\item Observed stellar positions and observed absolute velocities where $1/3$ of the stars are guaranteed to have positive radial velocities and dominant radial motions with respect to the mean position and velocity of the $N$ nearest neighbours, introducing both a local net expansion and local radial anisotropy (case III).
	\item Random stellar positions and velocities drawn from a four-dimensional Gaussian distribution with dispersions matched to those of each observed association (case IV).
	\item Random stellar positions and velocities drawn from a four-dimensional Gaussian distribution with dispersions matched to those of each observed association, where $1/3$ of the stars are guaranteed to have positive radial velocities, introducing a net expansion (case V).
	\item Random stellar positions and velocities drawn from a four-dimensional Gaussian with dispersions matched to those of each observed association, where $1/3$ of the stars are guaranteed to have positive radial velocities and dominant radial motions, introducing both a global net expansion and radial anisotropy (case VI).
\end{enumerate}
These models have been re-generated for the {\it Gaia}-DR2 data. The case I, II, and III models use the stellar positions and velocity magnitudes of the new DR2 sample. For the case II and case III models, the standard number of nearest neighbours used has been increased from $N=20$ to $N=100$, as typically the number of stars in each DR2 association is higher. Note that later in this work, we investigate the effect of varying the number of nearest neighbours used to generate the case II and case III models (see Section \ref{Nneighbours}). For the case IV, V, and VI models, we use values of the medians and standard deviations of the positions and velocities of the DR2 sample to generate the positions and velocity magnitudes of the model associations (see \citealt{Paper1} for details).

Together, these cases span a wide range of configurations, suitable for comparison to the observed kinematic diagnostics described in this paper. The case I and IV models represent randomised velocity fields with and without the observed positional substructure. The case II and III models represent expansion from multiple points spread throughout each association, with and without elevated radial anisotropy. The case V and VI models represent expansion from a single point at the centre of each association, with and without radial anisotropy.

\section{Quantifying the kinematics of OB associations}

\label{results_section}

In this work we quantify four key kinematic properties of the OB associations: the ratio between the number of sources moving away from and towards the association centre; the median radial velocity; the median of the radial velocities normalised by the tangential velocities; and the radial anisotropy parameter. In this context, radial velocity refers to the radial component of the velocity field in the plane of the sky with respect to the centre of each association. Each of these parameters is calculated from the two-dimensional velocity field in the plane of the sky such that the radial velocity component is defined with respect to the association centre and is perpendicular to the line-of-sight direction. In each of the first three diagnostics, associations in a state of expansion are expected to exhibit positive values. For the radial anisotropy parameter, positive values may be expected for a radially-anisotropic expansion (or contraction); however, as shown in \citet{Paper1}, a highly chaotic expansion from many originating points may not result in increased levels of radial-anisotropy. Nevertheless, these expanding velocity fields still exhibit increased radial velocities and fractions of stars in outward motions.

The cumulative distributions over the observed sample of 109 OB associations for each of these kinematic properties are presented in the first four panels of Fig. \ref{DR2_mainplot}. For each association, we calculate these quantities with respect to two different estimates of the association centres: one as the mean position and velocity of all stars in each association (solid black curves) and one using the mean position and velocity of only the OB-type stars in each association (solid red curves). It is assumed that the true centre of mass of a given association will lie between these two centre estimates as one is biased towards the rarest and most massive stars while the other is biased towards the far more numerous low-mass stars. 
For the most part, the results are insensitive to this choice. 
The calculated kinematic properties are listed for each observed association in Table \ref{app_tab2}. For simplicity, unless otherwise stated, where numbers are quoted within the text, they are drawn from the distributions calculated using the mean position and velocity of all stars in each association.

In order to test the impact of contamination, we randomly remove 10\% of the members of each association and repeat the measurement of each of the properties shown in Fig. \ref{DR2_mainplot}. Performing this process 100 times, we find that the results of this analysis are insensitive to the random removal of association members. Similarly, we test the dependence of the resulting distributions on the determination of the centre of each association by taking a weighted average position of all stars weighted by $G-$band flux and by $G-$band magnitude as simple proxies for mass. No variation is found beyond the stated uncertainties as a result of these additional tests.

\subsection{Positive-negative radial-velocity number ratios}

\begin{figure*}
	\begin{minipage}{170mm}
		\begin{center}
			\includegraphics[width=0.73\linewidth]{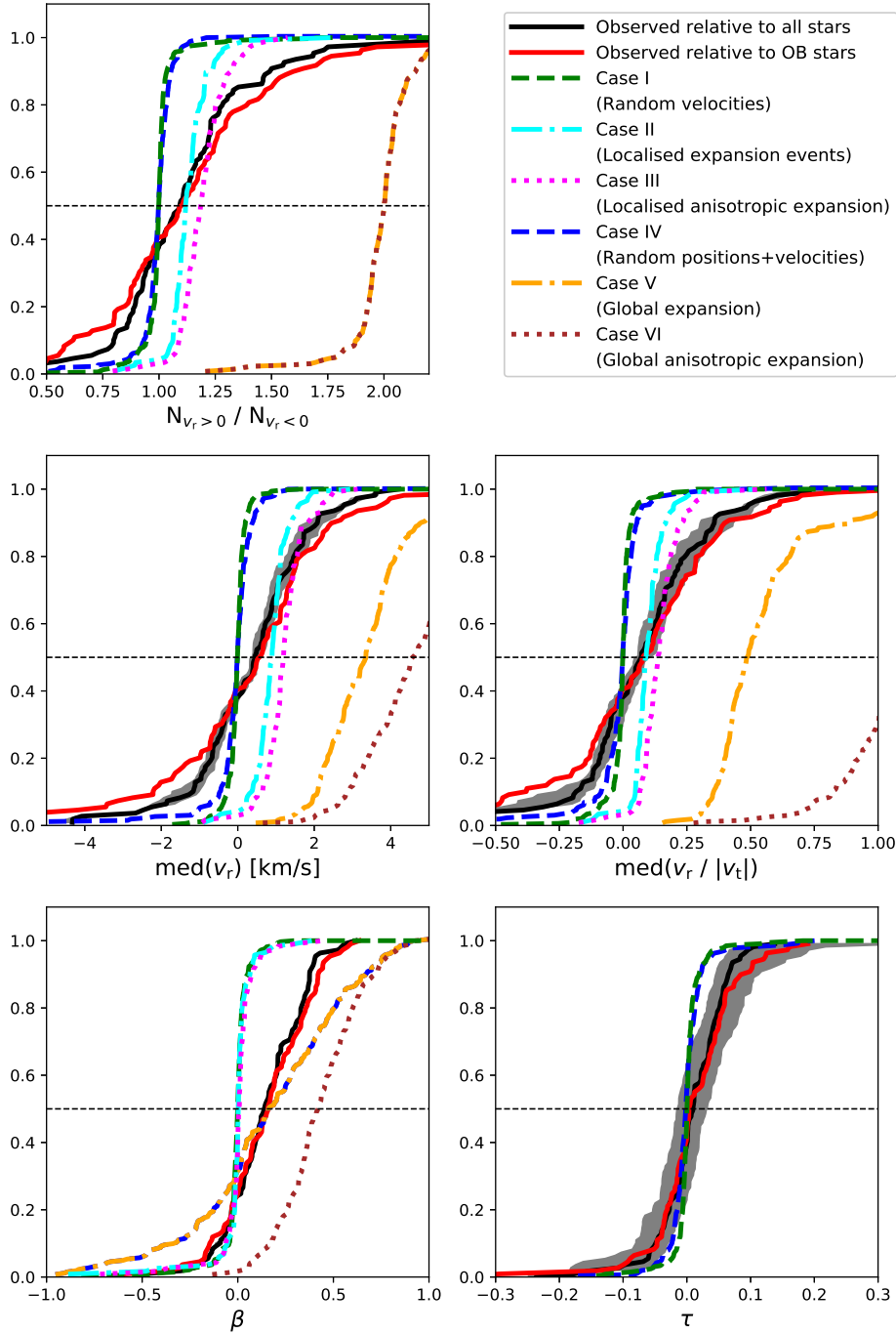}
\end{center}
\caption{\label{DR2_mainplot} Comparison of the observed OB associations with six models using various kinematic diagnostics (see Section \ref{results_section}). The observed distributions using the mean position and velocity of all stars in each association is shown in black, while the same distributions using the mean positions and velocities of only the OB-type stars in each association are shown in red. Uncertainties in the observed distribution relative to all stars are represented by the grey shaded area. Note that the uncertainties in $\beta$ are smaller than the thickness of the line shown. Also shown are the various model distributions with randomised velocity fields (dashed lines), expanding velocity fields (dash-dot lines), and radially-anisotropic velocity fields (dotted lines). The models that use the observed positions of the stars in each association are marked in green, cyan, and magenta (case I, case II, and case III) and those that use a random distribution of stellar positions are shown in blue, orange, and maroon (case IV, case V, and case VI). The ratio N$_{v_{\text{r}}>0}/$N$_{v_{\text{r}}<0}$ is shown in the upper-left panel, the median radial velocity in the middle-left panel and the median value of radial velocity normalised by tangential velocity in the middle-right panel. The lower-left panel shows the cumulative distributions for the radial anisotropy parameter ($\beta$) and the lower-right panel shows the observed (relative to the centre of all stars) and random (case IV) distributions for Kendall's $\tau$ measured for radial velocity against radius (see Section 3.4 for details). The horizontal black dashed line in each panel marks the median. The expanding model distributions are excluded from the lower-right panel, because any meaningful correlation between radius and radial velocity requires an evolution with time, yet our expanding velocity fields are not explicitly evolved and therefore do not differ from the randomised velocity fields. While only the globally expanding models are completely ruled out, none of these simple model distributions reproduce the observations over the entire parameter space of any of the presented diagnostics.
}
\end{minipage}
\end{figure*}

The upper-left panel of Fig. \ref{DR2_mainplot} shows the cumulative distribution over all observed OB associations for the ratio of the number of stars moving away from the centre of each association over the number of stars moving toward the centre (both relative to the centre of all stars and relative to the centre of only the OB-type stars). If the associations are undergoing a process of rapid and systematic expansion, it may be expected that the majority of stars should be moving away from the centre of each association and, within a monolithic formation scenario, one may expect this ratio to always be above one. 
With a median ratio of 1.07, a slight majority of stars do exhibit positive radial velocities in most, but certainly not all, associations. 
However, the excess in this ratio (with respect to unity) is relatively small and the ratio is less than 1.5 for more than 95\% of the associations, with a maximum measured value of 1.90. The ratio is lower than unity in the case of $\sim$1/3 of the associations with a minimum of 0.43.

The cumulative distributions for each of the model cases are also shown in the upper-left panel of Fig. \ref{DR2_mainplot}.
In \citet{Paper1}, we find that the observed distributions were best fit by randomised velocity fields. However, with the increased numbers of stars present in DR2, the randomised velocity field models (case I and IV) fall extremely close to 1.0 throughout the  distribution. The case IV model distribution (randomised positions and velocity fields) falls close to the observed distributions at ratios less than unity, as do the case II and III models that represent expansion from multiple points spread throughout each association. However, beyond this, the observed distributions diverge from the case II and III model distributions towards higher ratios. At no point are the observed distributions consistent with the singular expansion case V and VI models.

\subsection{Median velocity distributions}

The cumulative distributions of the median radial velocities of the observed sample  and the model associations are shown in the middle-left panel of Fig. \ref{DR2_mainplot}.  The two observed distributions diverge (but remain relatively close) in the upper portion of the distribution. This is likely due to the lower number of points determining the centre when using the OB star population compared to the entire sample for each association. As can be expected from the number ratios presented in the previous section, one third of the associations exhibit negative median radial velocities while two thirds exhibit positive median radial velocities. The median of the entire distribution lies at 0.4\,km\,s$^{-1}$\footnote{While perspective effects can cause erroneous detections of low-velocity expansions, this perspective expansion effect is dependent on distance \citep{vanLeeuwen2009}. We find no correlation between distances to associations and their median radial velocities (see Tables \ref{app_tab1} and \ref{app_tab2}). Therefore, it is unlikely that this figure results from the perspective expansion effect.} which while positive, does not immediately suggest a large-scale systematic expansion. Approximately 90\% of associations exhibit a median radial velocity of less than 1.5\,km\,s$^{-1}$, with a minimum and maximum observed median $v_{\text{r}}$ of $-$6.6\,km\,s$^{-1}$ and 2.6\,km\,s$^{-1}$, respectively. However, only two associations (associations 1 and 105) exhibit median radial velocities lower than $-2.6$\,km\,s$^{-1}$.

The case V and VI models that are representative of expansion from a single point at the centre of each association fall furthest from the observed distributions, and are completely inconsistent with the observed sample. 
The case I and IV distributions, representing randomised velocity fields consistently fall at lower velocities than the observed distributions in the positive radial velocity regime. The case IV distribution, representing models with randomised velocity fields and randomised stellar positions, is consistent with the observed distributions where the median velocity is less than zero (approximately one third of the observed associations).

As in \citet{Paper1}, none of the expanding model distributions fall close to the observed distributions across the entirety of the parameter space. However, the case II model distribution (multiple centres of isotropic expansion) is consistent with the observed distribution in the region of $1.0<v_{\text{r}}<1.3$\,km\,s$^{-1}$ when using the mean position and velocity of all stars in the associations as the centre of each association. 

The cumulative distributions of the median values of the radial velocities normalised by the tangential velocities (med$[v_{\text{r}}/|v_{\text{t}}|]$) are presented in the middle-right panel of Fig. \ref{DR2_mainplot}. The two observed cumulative distributions are consistent with one another within uncertainties. Again, one third of associations exhibit negative radial velocities, with the majority of the associations exhibiting positive values of med$(v_{\text{r}}/|v_{\text{t}}|)$. The median value is very close to zero (0.06) with minimum and maximum values of $-$0.78 and 0.42, respectively.

Again, none of the model distributions are completely consistent with the observed distributions but the models representing expansion from a single point (case V and case VI) fall at much higher values of med$(v_{\text{r}}/|v_{\text{t}}|)$ than the observed distributions and must therefore be ruled out.  The case II model distribution (multiple expansions) crosses the observed distributions in the rage $0.10 < v_{\text{r}} / |v_{\text{t}}| < 0.15$. The case III distribution (radially-anisotropic multiple expansions) is consistent with observations at values greater than 0.18.

\subsection{Radial anisotropy}

Elevated levels of radial anisotropy are a firm prediction of simulations of clusters undergoing gas-expulsion-driven expansion. This is expected to persist over several tens of Myr \citep{Baumgardt2007}; covering the entire lifetimes of the most massive stars in OB associations. 
We define the two-dimensional radial anisotropy parameter as:
\begin{equation}
\label{anieqn}
\beta = 1-\frac{<v_{\text{t}}^{2}>}{<v_{\text{r}}^{2}>} \text{.}
\end{equation}
This differs from the three-dimensional definition by omitting the factor of 2 in the denominator used to correct the double dimensionality of the tangential component in three dimensions.

The cumulative distribution of radial anisotropy parameters is shown in the lower-left panel of Fig. \ref{DR2_mainplot}. The majority ($>78$\%) of observed associations exhibit positive values of $\beta$, indicative of radially-anisotropic velocity fields. The median value of radial anisotropy parameter is $\beta = 0.11$, which while positive, is very low. Even the maximum measured value ($\beta$=0.40) remains low when considering that \citet{Baumgardt2007} predict values of $\beta > 0.5$ for expanding clusters after ten crossing times when at least one fifth of the mass is encompassed within the radius considered. Only 6\% of associations exhibit anisotropies of $\beta > 0.3$.

As for all of the previously discussed parameters, the radially-anisotropic singular expansion model distribution (case VI) is by far the least consistent with observations, exhibiting considerably more extreme levels of anisotropy than those observed. Therefore, this scenario can be conclusively ruled out as the origin of OB associations. However, it is also clear that none of the model distributions provide a reasonable fit to the observed distributions. The case IV and V model distributions are entirely inconsistent with observations, exhibiting a very different profile, despite crossing the observed distributions at the approximate median. The case I and II distributions fall closer to the observed distributions but, again, simply cross the distributions and are inconsistent with observations. Therefore, while the singularly monolithic expansion scenario is completely ruled out as the dominant means by which OB associations are formed, none of the model cases presented are consistent with the observed kinematic properties of OB associations. In Section 4 we investigate which elements are necessary in order to reproduce the observed cumulative kinematic properties of OB associations.

\subsection{Radial velocity sorting}

\label{radvelsortSec}

Under a monolithic formation scenario for OB associations, stars found furthest from the centre of the association, must necessarily exhibit a higher (more positive) radial velocity. This also holds true within a multiply monolithic model, albeit in that case we can expect a great deal of scatter around the underlying relation. This radial velocity sorting is therefore a key behaviour to look for when investigating the dynamical history and evolution of young clusters and associations.

\begin{figure*}
	\begin{minipage}{170mm}
		\begin{center}
			\includegraphics[width=\textwidth]{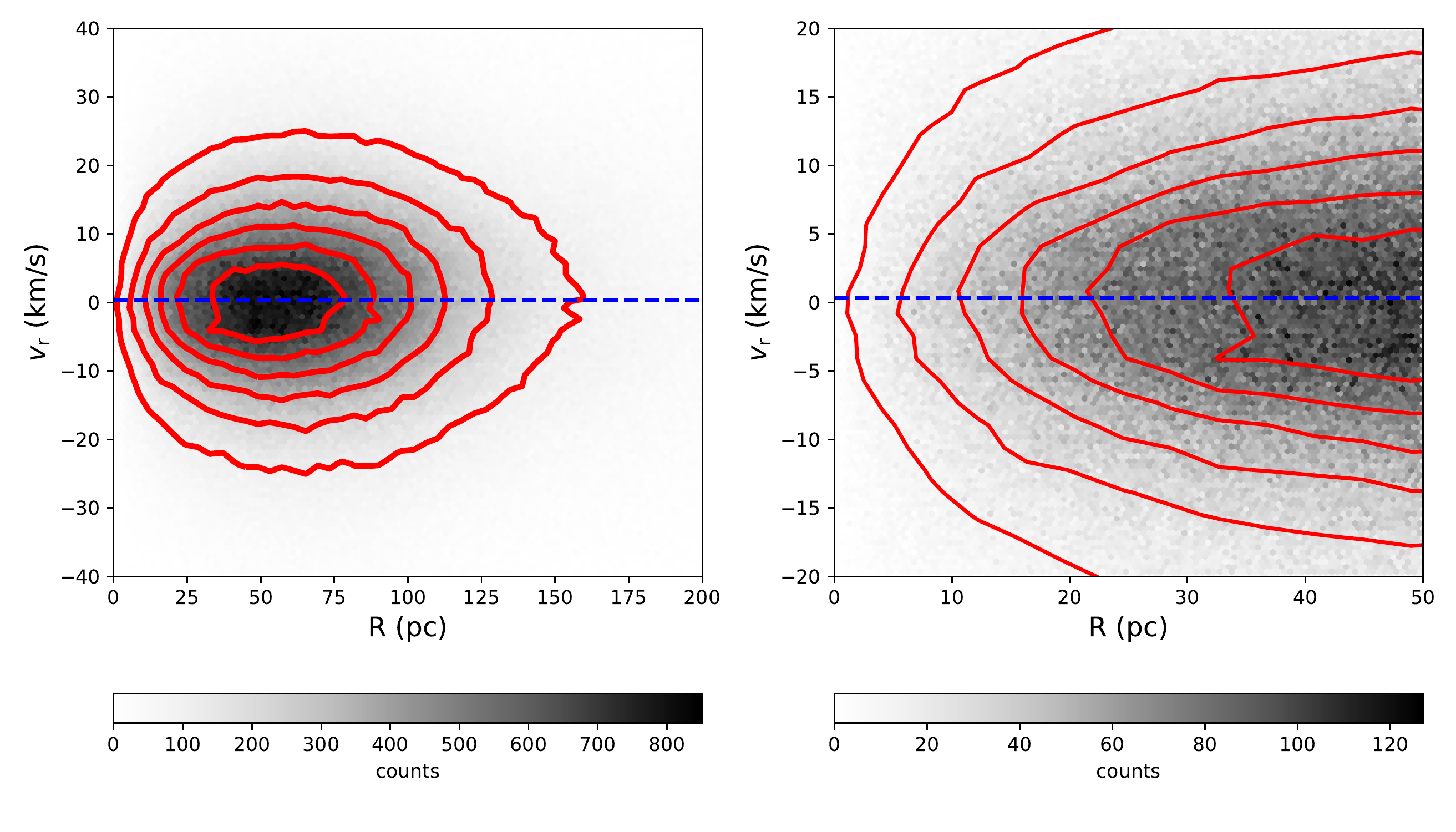}
		\end{center}
		\caption{\label{vr_vt_fig} Left panel: radial velocity versus radius from association centre for all stars in the sample. Red contours represent the number density and the blue dashed line marks the median radial velocity over the entire sample (0.0007\,km\,s$^{-1}$). Right panel: Same as the left panel but restricted to R$<$50\,pc and $-20<v_{\text{r}}<20$\,km\,s$^{-1}$.}
	\end{minipage}
\end{figure*}
\begin{figure*}
	\begin{minipage}{170mm}
	    \begin{center}
			\includegraphics[width=0.49\linewidth]{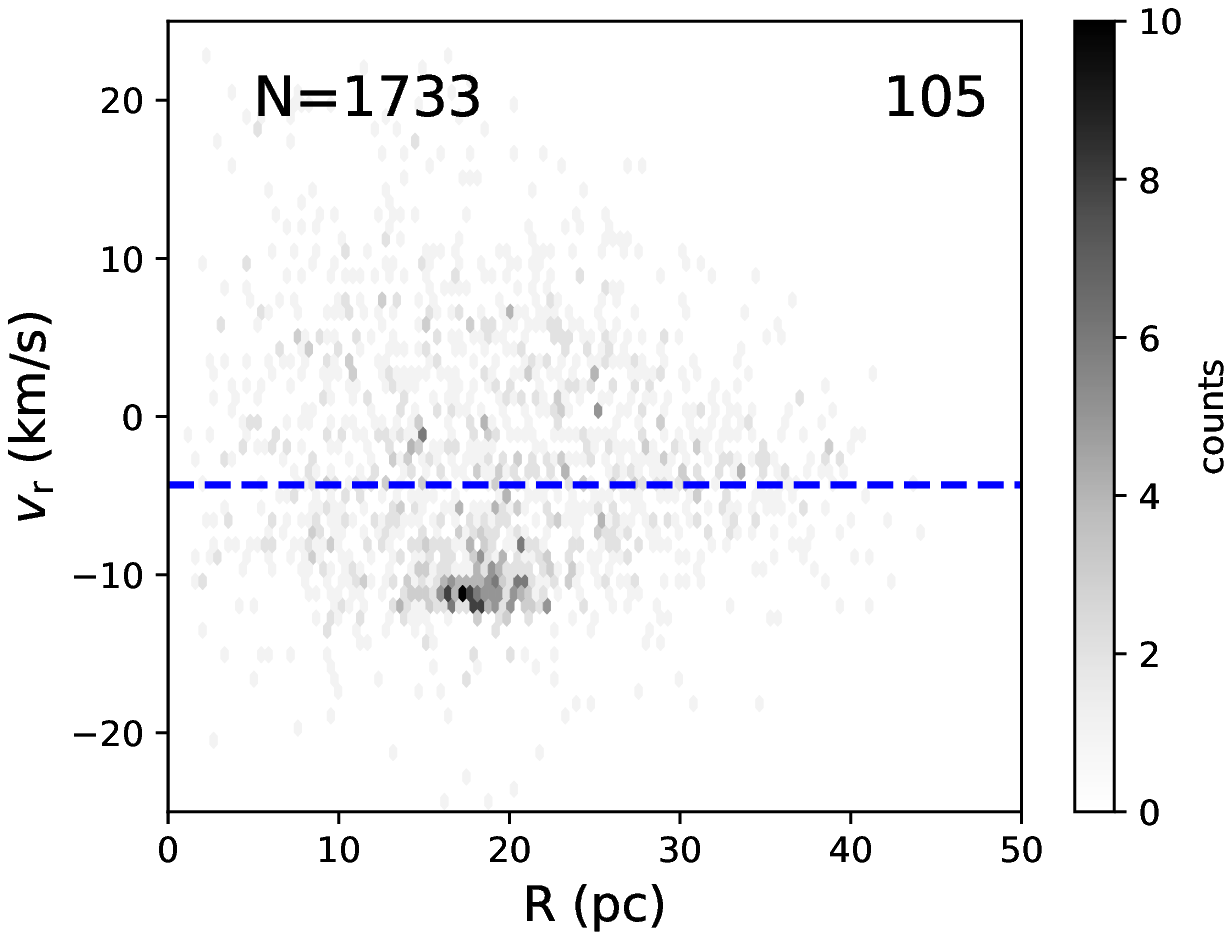}
			\includegraphics[width=0.49\linewidth]{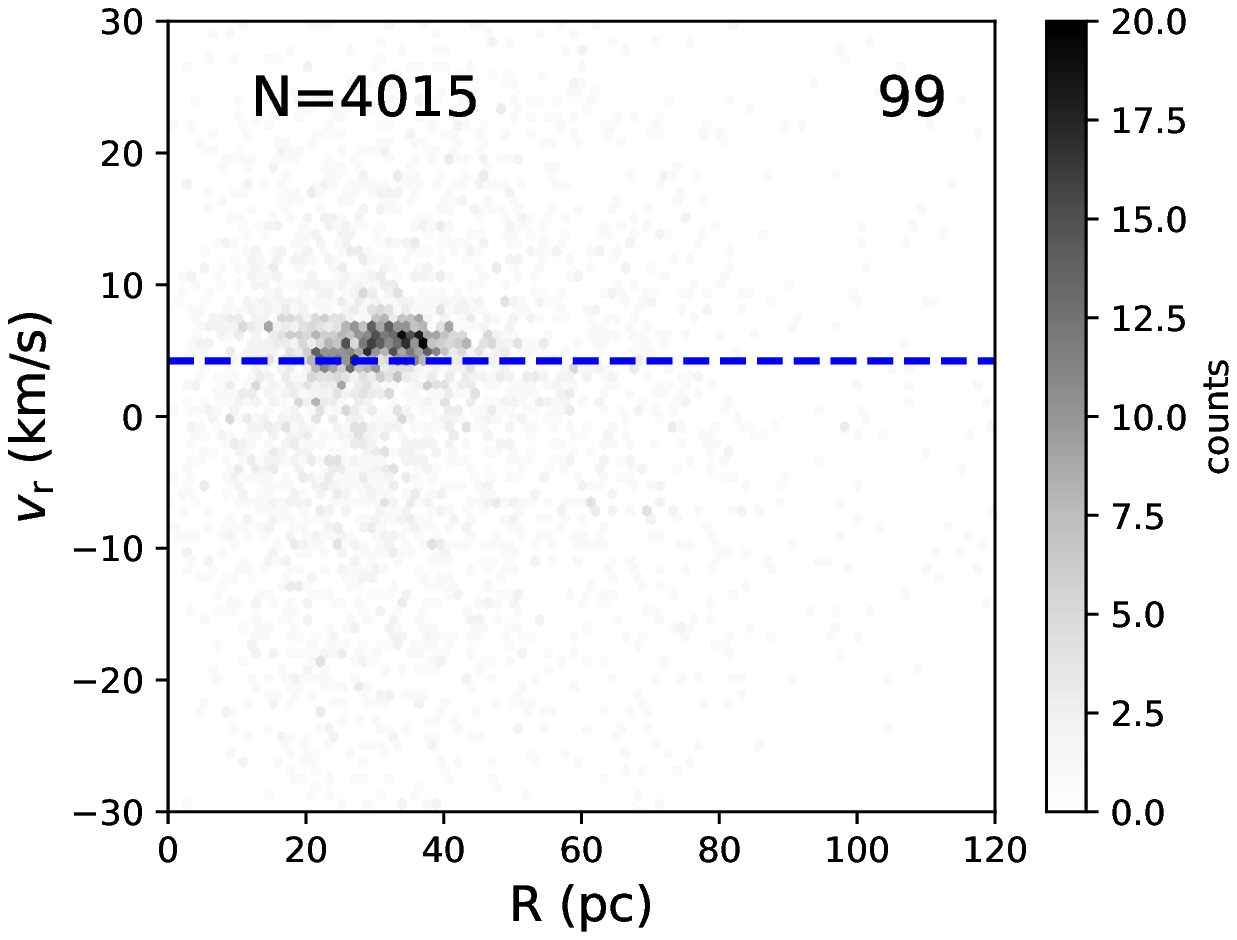}
			\includegraphics[width=0.49\linewidth]{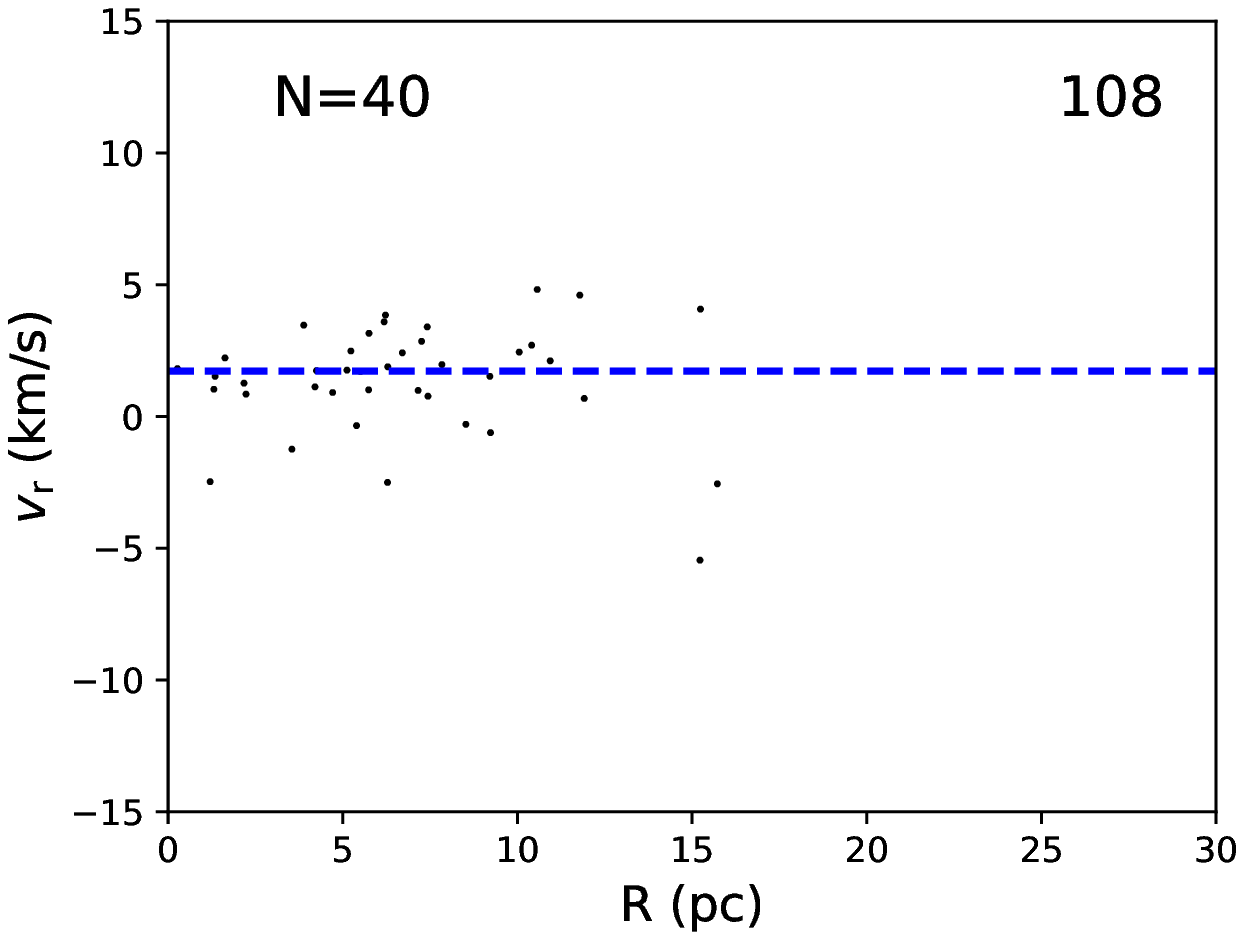}
			\includegraphics[width=0.49\linewidth]{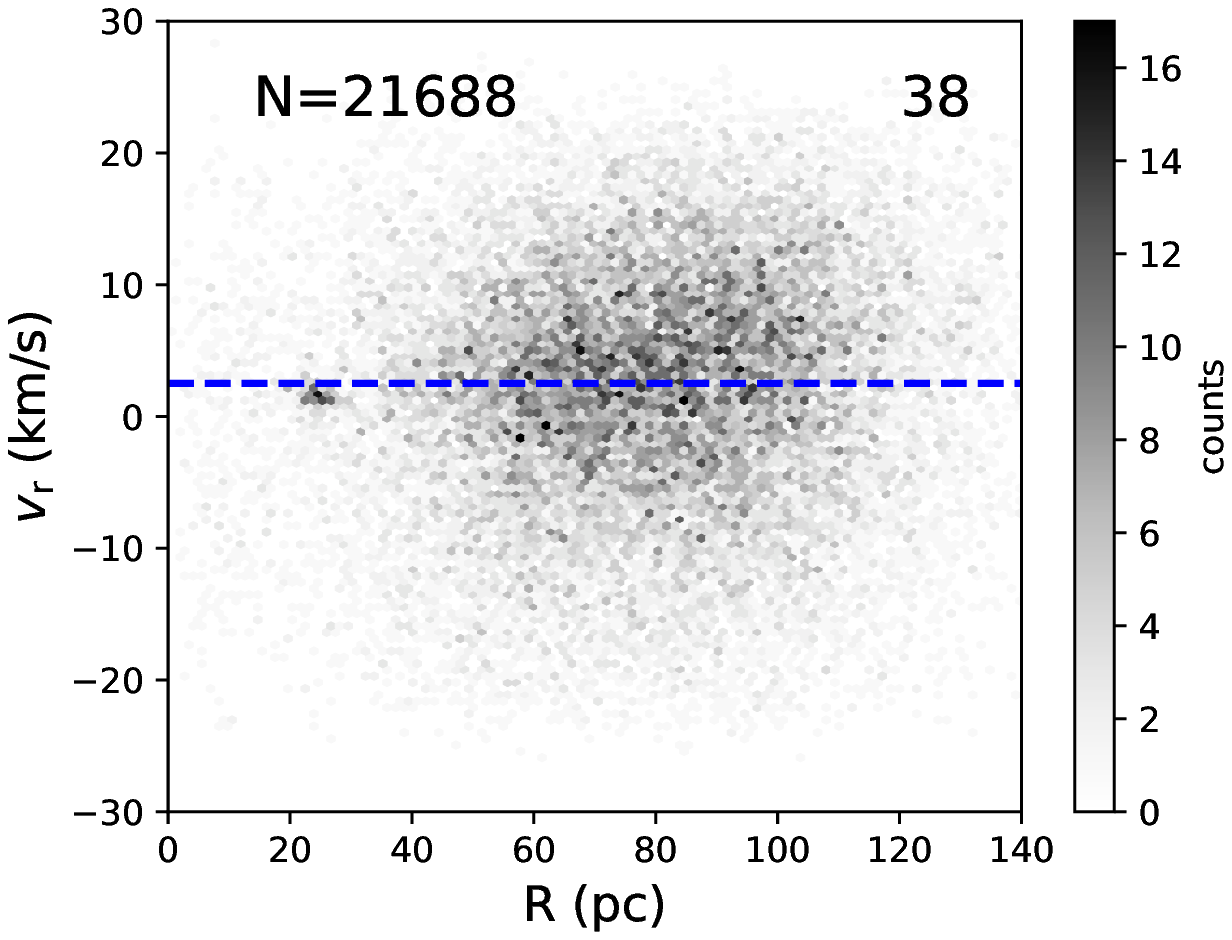}
		\end{center}
		\caption{\label{example_Vr_R} Here we show individual $v_{\text{r}}$ versus R diagrams for four randomly selected associations (associations 105,99,108, and 38). The lower-left panel uses discrete points for each star due to the low number of stars assigned to association 108 while the other figures make use of hexbins shaded according to density. Note that there is no significant correlation between radial velocity and radius in any of the associations, regardless of radial extent, number of stars, or the presence of velocity substructure. This indicates that these associations could not have been formed by the expansion of compact clusters.}
	\end{minipage}
\end{figure*}

In Fig. \ref{vr_vt_fig}, radial velocity is shown against radius for every star in the sample. There is no correlation between radial velocity and radius over the entire sample. However, it is possible that while no correlation is present in the sample as a whole, correlations may be present within individual associations that are masked by the sheer number of stars shown in Fig. \ref{vr_vt_fig}.
In Fig. \ref{example_Vr_R}, we present four examples of $v_{\text{r}}$ versus radius for individual OB associations selected at random. Similar figures have been produced for every association in the sample, and the randomly selected sample shown here is representative of the range of velocities and radii in the sample as a whole. For individual OB associations, as for the entire sample, no correlation is evident.

We quantify the degree to which correlations between radial velocity and radius are present by calculating the Kendall's $\tau$ rank-correlation coefficient between these quantities for each association in our sample. Kendall's $\tau$ is defined as:
\begin{equation}
    \tau = \frac{n_{c}-n_{d}}{N_{\ast}(N_{\ast}-1)/2}
\end{equation}
 where $n_{c}$ is the number of concordant pairs of data (such that within a pair, where R increases, so does $v_{r}$ and vice versa), $n_{d}$ is the number of discordant pairs and $N_{\ast}$ is the total number of stars in the sample. As such, in the case of perfectly correlated data $\tau = 1$ and in the case of perfectly anti-correlated data $\tau = -1$, with perfectly uncorrelated data yielding $\tau = 0$. The cumulative distribution of $\tau$ over all associations is shown in the lower-right panel of Fig. \ref{DR2_mainplot}. The median and mean values of $\tau$ are both approximately zero, indicating that there is no correlation between radial velocity and radius. The maximum measured value of  $\tau$ is $\sim$0.1, indicating that any potential correlations are extremely weak at best. Also shown in the same panel are the cumulative distributions of $\tau$ over the randomised velocity field (case I and IV) models. The model cases representing expanding velocity fields (cases II, III, V, and VI) are not shown because developing a relationship between $v_{\text{r}}$ and $R$ depends on the evolution of a structure with time. As the expansion-like velocity fields are inserted into the models in an instantaneous fashion, no such relation between radial velocity and radius can be expected. At all values of $\tau$, the observed distribution falls between the case I and case IV model distributions. There is therefore no evidence for radial velocity sorting over the large-scale structure of entire OB associations. Therefore, it seems extremely unlikely that the large-scale structure of stellar associations is arrived at via the expansion of clusters. Note that while dynamical mixing may remove the traces of radial velocity sorting over time, the elevated levels of anisotropy suggest that associations have not yet undergone sufficient mixing to remove traces of expansion events.

\section{Improved model associations}

\label{improved_models}

We now explore how the model associations used in Section \ref{results_section} can be improved in order to better describe the observed kinematic properties of OB associations. We begin by varying the number of nearest neighbours that define the centres of expansion from which stars are directed away from in the generation of the case II (multiple centres of expansion) and case III (multiple centres of anisotropic) model associations. Next we investigate the impact of introducing substructure in velocity space in addition to retaining the original positional substructure of the observed associations. Finally, we introduce small-scale expansion events with an age spread into the model associations with substructure in position and velocity space. These new models represent a step towards a more sophisticated interpretation of the available kinematic diagnostics in this work and allow us to investigate the effects of velocity substructure on the bulk kinematic properties of the associations, as well as the influence of small-scale expansions of substructures within associations.

\subsection{Varying the number of nearest neighbours}

\label{Nneighbours}

It is clear from the derived kinematic properties above that one of the key results of \citet{Paper1}, i.e. that the velocity fields of nearby OB associations are fully consistent with randomised velocity fields, is not applicable to the {\it Gaia}-DR2 sample. Meanwhile, the new observed distributions are completely inconsistent with singularly expanding model cases, but also not consistent with the multiple expansion model cases using 100 nearest neighbours to define the points from which stars are expanding. In order to quantify the effects of changing the nmber of nearest neighbours, we construct grids of case II and case III models. These new case II and III models use the same constant proportion of stars that are forced into expanding motions (33\%) and vary the number of nearest neighbours used to determine the local centre from which each of the \textquoteleft expanding stars\textquoteright are moving away from. Three model associations are generated for each observed association (109 associations), for each value of nearest neighbours (10 values between 5 and 3200), resulting in 3270 case II and 3270 case III model associations.

\begin{figure*}
\begin{minipage}{170mm}
	\begin{center}
	\includegraphics[width=1.0\linewidth]{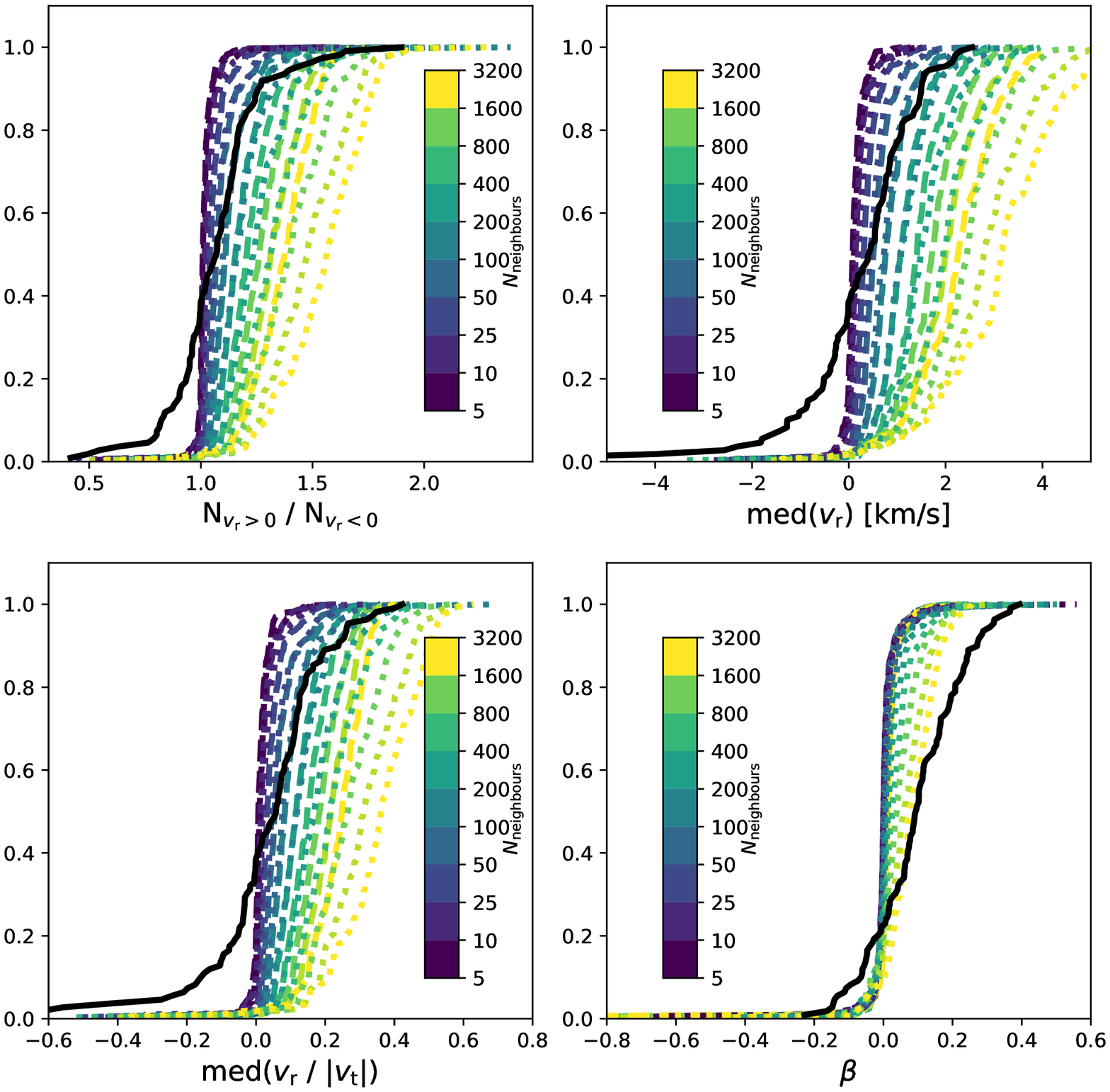}
	\end{center}
	\caption{\label{gridfigs} Cumulative distributions for four kinematic properties (N$_{v_{\text{r}}>0}/$N$_{v_{\text{r}}<0}$,$v_{\text{r}}$,$v_{\text{r}}/|v_{\text{t}}|$,$\beta$) of the case II (dashed curves) and case III (dotted curves) grids of models with varying numbers of nearest neighbours from 5 to 3200. Also plotted for reference is the observed distribution for each parameter, relative to the centre of all stars in each association (solid black curve). While many of the model distributions cross the observed distributions, no individual model can be considered consistent with observations across the entire parameter space.}
\end{minipage}
\end{figure*}

In Fig. \ref{gridfigs} we present the cumulative distributions for the four kinematic parameters shown in Fig. \ref{DR2_mainplot}, this time showing the grid of case II and III models using varying numbers of nearest neighbours to generate the model associations. Also shown is the observed distribution, using the mean position and velocity of all stars in each association to define each association centre. It is clear that a single number of nearest neighbours is insufficient to describe the cumulative distributions of the observed sample as the observed distribution crosses the parameter space of a significant proportion of the models. 

The case III models consistently exhibit higher values than the case II models for all values of $N_{\text{neighbours}}$. This is to be expected given that the radially-anisotropic models will always exhibit higher levels of radial motions compared to the case II models. For all four kinematic properties, there is an excess of observed associations with negative radial velocities and radial-anisotropies.  While a single model cannot fit the entire observed distribution, it is clear that the case II models are completely unable to reproduce the observed levels of elevated radial anisotropy. Any formation scenario based on the expansion of clusters must therefore include radially-anisotropic expansion.

\begin{figure}
    \centering
    \vspace{5mm}
    \includegraphics[width=0.95\linewidth]{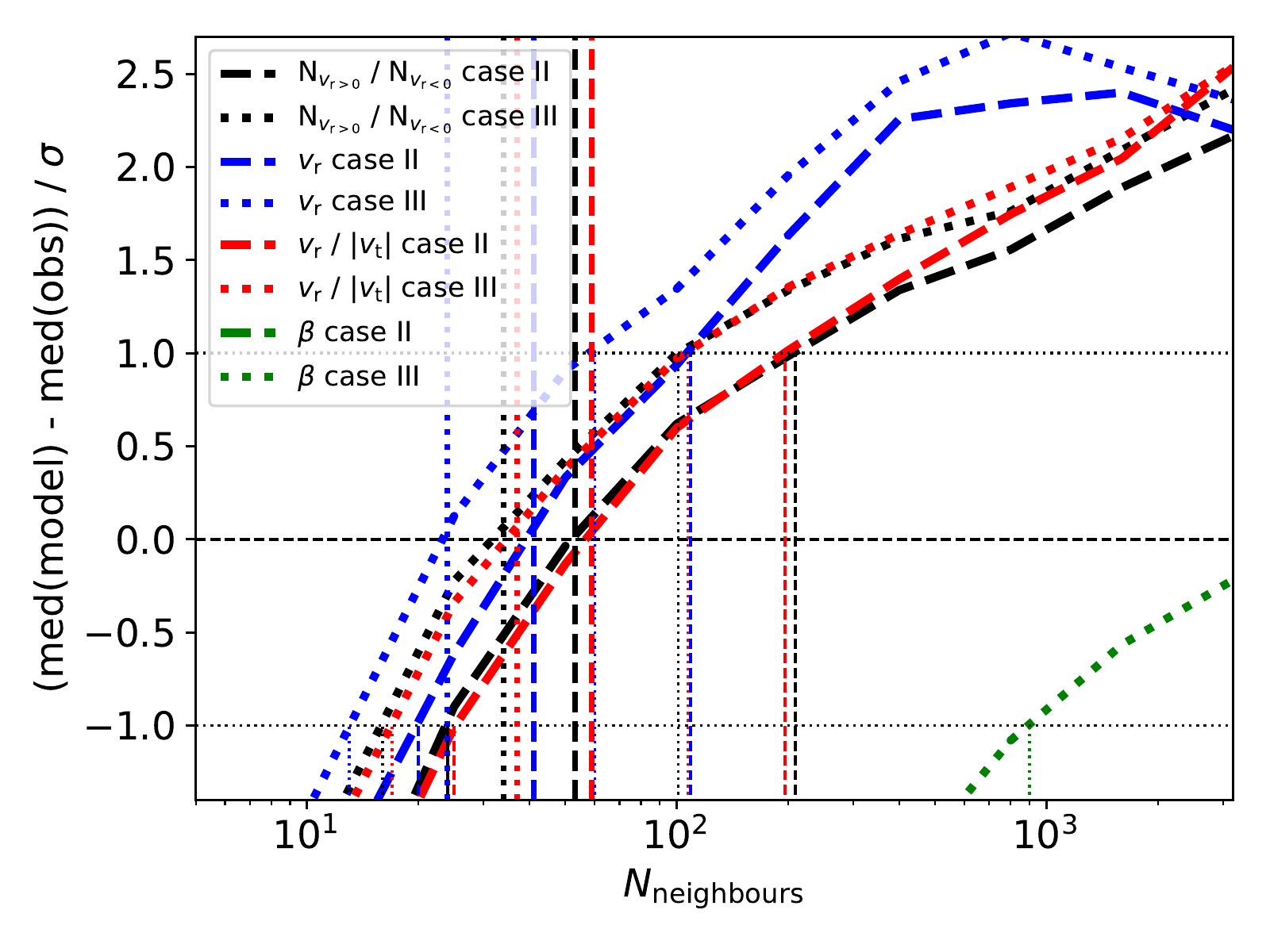}
    \caption{ A comparison between the number of nearest neighbours (used to define local centres of expansion) required when generating case II and III models to reproduce the median observed values for each kinematic diagnostic (see Section \ref{Nneighbours}). The difference between the median model value and the median observed value, normalised by the standard deviation of each distribution ($\sigma$) is shown versus the number of nearest neighbours used to generate the models. Note that the curve for the case II models and $\beta$ does not appear in the displayed range because the case II models are completely inconsistent with the observed levels of radial anisotropy. The vertical lines mark the number of nearest neighbours where each distribution has a difference of zero (thick lines),\, as well as $+1\sigma$ and $-1\sigma$ (thin lines). The number of nearest neighbours required to reproduce the median of the observed distribution differs for each diagnostic. In particular, the elevated levels of radial anisotropy cannot be reproduced by the same models that reproduce the observed median radial velocities. This implies that a level of sophistication beyond position substructure with simple localised expansion is required to reproduce the observed kinematics.}
    \label{fig:multi_grid_diff}
\end{figure}
In order to determine which model (on average) best describes the observations, we take the median of the each model distribution and subtract the median of the observed distribution. This is then normalised by the standard deviation of the model distribution. This is performed for each number of nearest neighbours and each of the four kinematic parameters (as shown in Fig. \ref{gridfigs}) and shown against the number of nearest neighbours in Fig. \ref{fig:multi_grid_diff}. The curve for case II and $\beta$ does not appear in Fig. \ref{fig:multi_grid_diff} because the median observed value exceeds the median radial anisotropy for case II models of any number of nearest neighbours. The number of nearest neighbours that yields a negligible difference between the medians of the model distribution and observed distribution is estimated by calculating the point at which each curve in Fig. \ref{fig:multi_grid_diff} crosses zero. The number of nearest neighbours this results in is presented in Table \ref{n_neighbours_xtab}. The number of nearest neighbours for the radial anisotropy parameter and the case II models is omitted as the case II models cannot recreate the observed levels of radial anisotropy. The number of nearest neighbours required to reproduce the median value of $\beta$ with case III models has not been calculated but must exceed the maximum value used in the grid of 3200.

The case III models consistently require a lower number of nearest neighbours than the case II models to reproduce the median observed value. 
The median number ratio (N$_{v_{\text{r}}>0}/$N$_{v_{\text{r}}<0}$) is best reproduced with 53 nearest neighbours in the case II models and 34 with the case III models. The median radial velocity parameters require similarly low (41 and 59 for case II, 24 and 37 for case III) values of nearest neighbours, indicating that a large number of local expansion centres is necessary to reproduce the median values of the observed radial velocity distributions.
However, it is impossible to reproduce the observed levels of radial anisotropy with multiple non-anisotropic expansions (case II) and the case III models require >3200 nearest neighbours to match the observed median. This is at least two orders of magnitude higher than the number of nearest neighbours needed for the other expansion metrics used here and exceeds the total number of stars in some of the associations. It would be possible to adjust the case III models by increasing the radial anisotropy for each star that is forced into an expanding motion, but this would also necessarily have the effect of increasing the radial velocity. This would therefore decrease the number of nearest neighbours required to best match the observed radial velocity properties, and have little or no effect on the discrepancy between the number of nearest neighbours required to reproduce the various kinematic properties.

\begin{table}
\caption{The number of nearest neighbours for which the difference between the median of the observed distribution and the median of the model distribution is zero. The quoted uncertainties are drawn from the point at which the distributions shown in Fig. \ref{fig:multi_grid_diff} cross the +1$\sigma$ and $-1\sigma$ lines. Note that this is only relevant for the case II and case III models and that this value cannot be determined for  $\beta$ using the case II model distributions, as they do not exhibit a significant excess in $\beta$. For the case III models, the median of the model distribution and observed distribution are equal at a number of nearest neighbours greater than the maximum number considered here. The quoted number in the table below is drawn from the intersection of the distribution shown in Fig. \ref{fig:multi_grid_diff} with $-1\sigma$.}
    \centering
    \begin{tabular}{l c c}
    \hline
    Parameter & case II & case III \\
    \hline
        N$_{v_{\text{r}>0}}/$N$_{v_{\text{r}<0}}$ & 53$\substack{+156\\-29}$ & 34$\substack{+67\\-18}$ \\
        $v_{\text{r}}$  & 41$\substack{+68\\-21}$ & 24$\substack{+36\\-11}$ \\
        $v_{\text{r}}/|v_{\text{t}}|$ & 59$\substack{+137\\-34}$ & 37$\substack{+70\\-20}$ \\
        $\beta$ & \null & $>$900 \\
    \hline
    \end{tabular}
    \label{n_neighbours_xtab}
\end{table}

The large discrepancy between the number of nearest neighbours required to reproduce the median observed kinematic properties of OB associations suggests that a simple model of multiple, radially-anisotropic expansions is unable to reproduce the kinematic properties of OB associations. Moreover, the inability of any of these models to reproduce the shape of the observed distributions, or the observed associations with negative median radial velocities indicates that models of this form are not capable of accurately representing the complete kinematic behaviour of stellar associations.

\subsection{Models with velocity substructure}

\label{velsubstructure}

While we include positional substructure in the class I, II, and III models, stellar associations are also likely to exhibit significant levels of substructure in velocity space. Therefore, in this section, we introduce velocity substructure into the case I models that consist of a randomised velocity field and the original stellar positions of the observed OB associations, therefore including any positional substructure present in the observations.

\begin{figure*}
	\includegraphics[width=\linewidth]{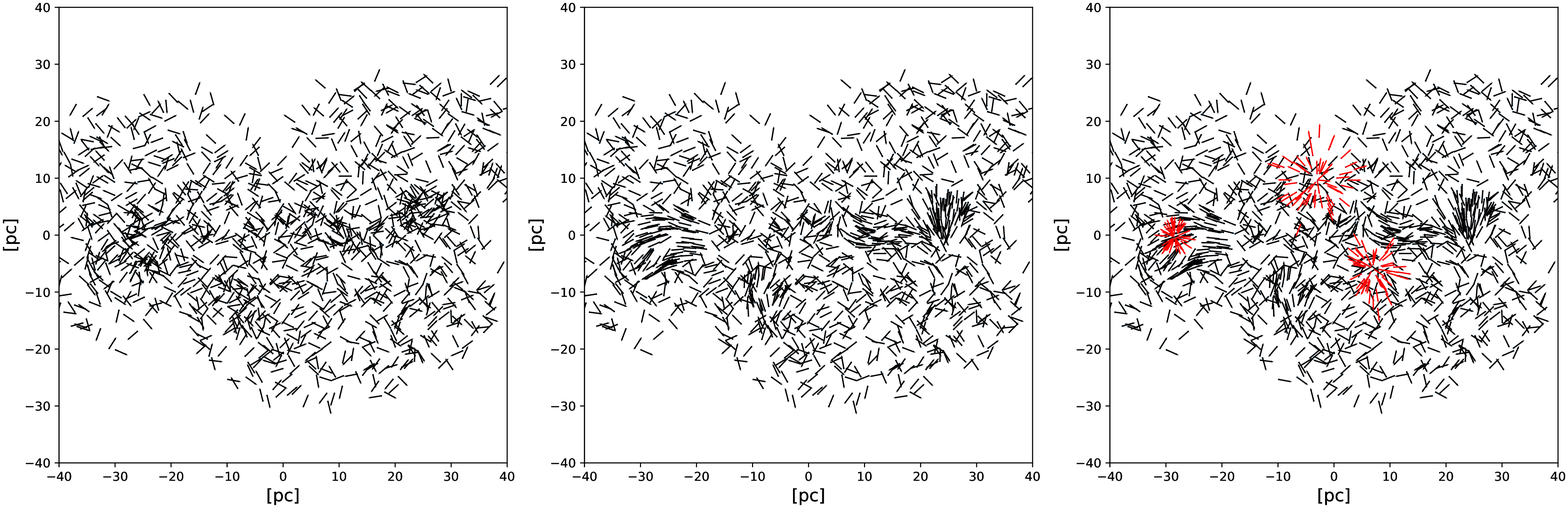}
	\caption{\label{subvel_example} Velocity vector maps for model associations based on association 96. Left: Case I model that consists of the observed positional structure and a randomised velocity field. Centre: Case I model plus velocity substructure. Right: Case I model plus velocity substructure with additional expanding substructures added. The additional stars that make up the expanding clusters added to the right-hand panel are shown in red. All velocity vectors are normalised and are indicative only of direction. Note that the directions of motion of the two prominent overdensities in the left hand panel align in the centre panel.}
\end{figure*}

Velocity substructure is introduced into the models by ensuring that stars that lie close together in positional space also exhibit similar velocities. This is done using a grid over two free parameters: the scale over which stars are considered to be nearby to one another, and the number of nearest stars used to form a nearby group. If a star falls within the threshold distance ($d_{\text{threshold}}$) of a threshold number of stars ($n_{\text{threshold}}$), the velocity of that star is modified such that it moves in the mean direction of motion of those $n_{\text{threshold}}$ nearest neighbours, plus a 5\% level of noise drawn at random from a Gaussian distribution. The threshold number of stars is set as:
\begin{equation}
    n_{\text{threshold}} = c_{\text{n}}N_{\ast}
\end{equation}
where $c_{\text{n}}$ is the fraction of the total number of stars to use as the group of nearest stars and $N_{\ast}$ is the total number of stars in each association. The threshold distance is similarly set as:
\begin{equation}
    d_{\text{threshold}} = c_{\text{d}}\bar{d} \text{,}
\end{equation}
where $c_{\text{d}}$ is the threshold scale and $\bar{d}$ is the mean distance between stars in each association. We generate a grid of substructured models using four values of $c_{\text{n}}$ (0.01, 0.02, 0.05, 0.10) and five values of $c_{\text{d}}$ (0.05, 0.1, 0.2, 0.3, 0.5). Figure \ref{subvel_example} shows a normal case I model with a randomised field based on association 96 (left panel), as well as the model with added velocity substructure, obtained as described above (centre panel). Association 96 was chosen as an example because overdensities are clearly visible in position space and the relatively low number of members (818) assigned to this association allows for the effects of this process to be  visible to the naked eye.

\begin{figure*}
	\begin{minipage}{170mm}
		\begin{center}
			\includegraphics[width=0.8\linewidth]{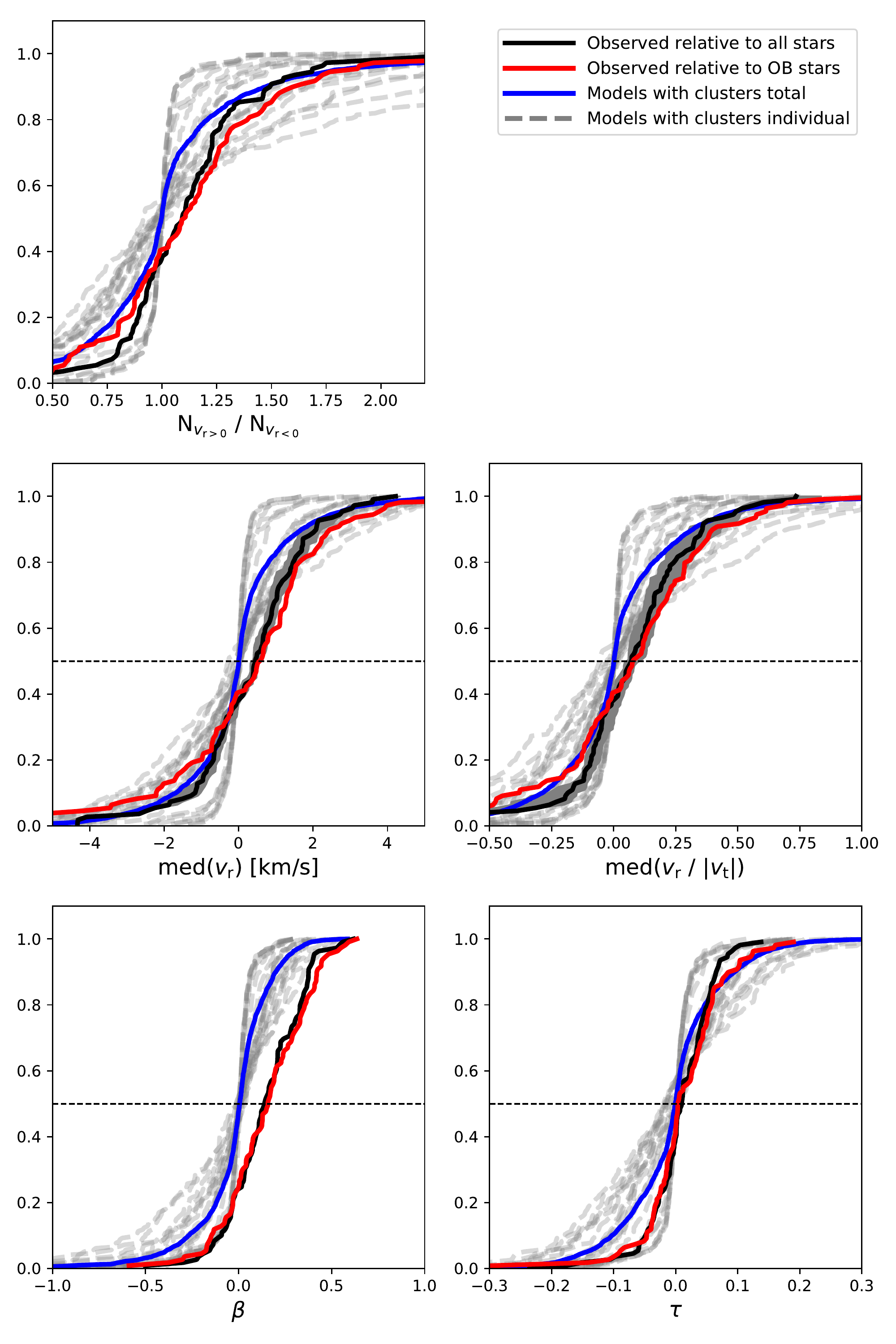}
\end{center}
\caption{\label{6panel_subvel} Cumulative distributions of each of the five kinematic parameters for each set of model parameters described in Section \ref{velsubstructure} are shown in the grey dashed lines with the total combined distribution over all of the models with velocity substructure is shown in the blue solid curve. The cumulative distributions of the observed sample of OB associations are shown in the black and red solid lines. Uncertainties in the observed distribution have been omitted for clarity but can be found in Fig. \ref{DR2_mainplot}. The ratio N$_{v_{\text{r}}>0}/$N$_{v_{\text{r}}<0}$ is shown in the upper-left panel, the median radial velocity in the middle-left panel and the median value of radial velocity normalised by tangential velocity in the middle-right panel. The lower-left panel shows the cumulative distributions for the radial anisotropy parameter ($\beta$) and the lower-right panel shows the distributions for Kendall's $\tau$ measured for radial velocity against radius using the centre of all stars in each association. These new model distributions produce a much broader distribution for each of the kinematic diagnostics than the purely randomised velocity field models of Section \ref{results_section}. This is much more reminiscent of the observed distributions but there remains an observed excess compared to the model distributions in number ratio, median radial-velocities, and radial-anisotropy. This implies that a further level of sophistication (i.e.\ beyond velocity substructure) needs to be added to the models to reproduce the observed kinematics.
}
\end{minipage}
\end{figure*}

These models are then analysed in the same manner as the observations and all other model distributions in this work. The resulting cumulative distributions for each combination of $c_{\text{n}}$ and $c_{\text{d}}$ are shown in grey in Figure \ref{6panel_subvel}, with the total combined cumulative distribution over all of the models in the grid shown in blue. Note that no single combination of $c_{\text{n}}$ and $c_{\text{d}}$ is equivalent to the total cumulative distribution over all model associations. For all parameters except Kendall's $\tau$ for $v_{\text{r}}$ versus radius, the observed cumulative distributions exhibit a small yet statistically significant excess towards the expanding regime compared to the model distributions, indicating that the inclusion of velocity substructure is not sufficient to explain the observed kinematic properties of OB associations. However, the shape of the total cumulative distribution over all substructured models is much more similar to the observed distributions than in all previous models, suggesting that kinematic substructure is a necessity in order to reproduce the cumulative kinematic properties of the observed associations.

\subsection{Localised expansion events}

\label{popclusters}

While the lack of velocity sorting within OB associations (see Section \ref{radvelsortSec}) is certainly indicative that the expansion of more compact clusters does not set the overall size of associations, it is certainly likely that some stars within OB associations form in highly-clustered substructures \citep[e.g.][]{Gouliermis2018} and these substructures may undergo localised expansion events. With that in mind, we produce a new set of models in which a small number of localised expansion events are placed within each of the above substructured model associations. This approach differs from the case II and III models (that also represent localised expansion events) in two key ways. First, these models include substructure in both position and velocity as inherited from the models of Section \ref{velsubstructure}, whereas the case II and III models include substructure in positional space only. Second, each localised expansion event in the new models is a discrete event from a specific location and tracked over a period of time, while the case II and III models represent an expansion from an indeterminate number of points spread throughout each association.

The number of clusters assigned to each association is generated at random from a Gaussian distribution. A mean of 5 clusters are injected into each association with a standard deviation of 1 and minimum and maximum numbers of clusters of 2 and 10, respectively. The number of stars that make up this new clustered population are defined as a multiple of the number of stars in the original model, $n_{\text{clust}} = c_{\text{frac}}N_{\ast}$. These additional stars are then distributed evenly between the new clusters. Each cluster is assigned a randomly determined age, again drawn from a Gaussian distribution with a minimum age of zero. Each star is assigned velocity components in both the X and Y directions, drawn from a normal distribution based on the mean and standard deviation of the velocities of the stars in the model association. The final stellar positions of the newly added stars are then calculated using the cluster age, assuming a constant velocity. The addition of these clusters therefore represents a small number of maximally anisotropic expansion events within each substructured model association. We also include the velocity substructure considered in Section \ref{velsubstructure}.

We generate these new models across a three parameter grid varying the fraction of the number of stars to use to generate clustered population numbers ($c_{\text{frac}}$), the mean age of clusters, and the standard deviation of cluster age. Five values of the clustered population fraction (0.001, 0.002, 0.005, 0.01, 0.02), six values for the mean age of the clusters in each association (0, 0.5, 1.0, 2.0, 4.0, 8.0\,Myr), and six values of the cluster age dispersion (0.05, 0.1, 0.2, 0.5, 1.0, 2.0\,Myr) are used. This, in combination with the 6540 models with substructured velocity fields, yields a total grid size of $\sim$1.2 million model associations spanning 180 combinations of cluster initialisation parameters. All stars within each cluster have the same age. An example of the resulting models is shown in the right-hand panel of Fig. \ref{subvel_example} with the newly added expanding clusters marked in red. The mean cluster age used in this example is 0.5 with an age dispersion of 0.1, so the centre of each expansion remains clearly visible.

\begin{figure*}
	\begin{minipage}{170mm}
		\begin{center}
			\includegraphics[width=0.78\linewidth]{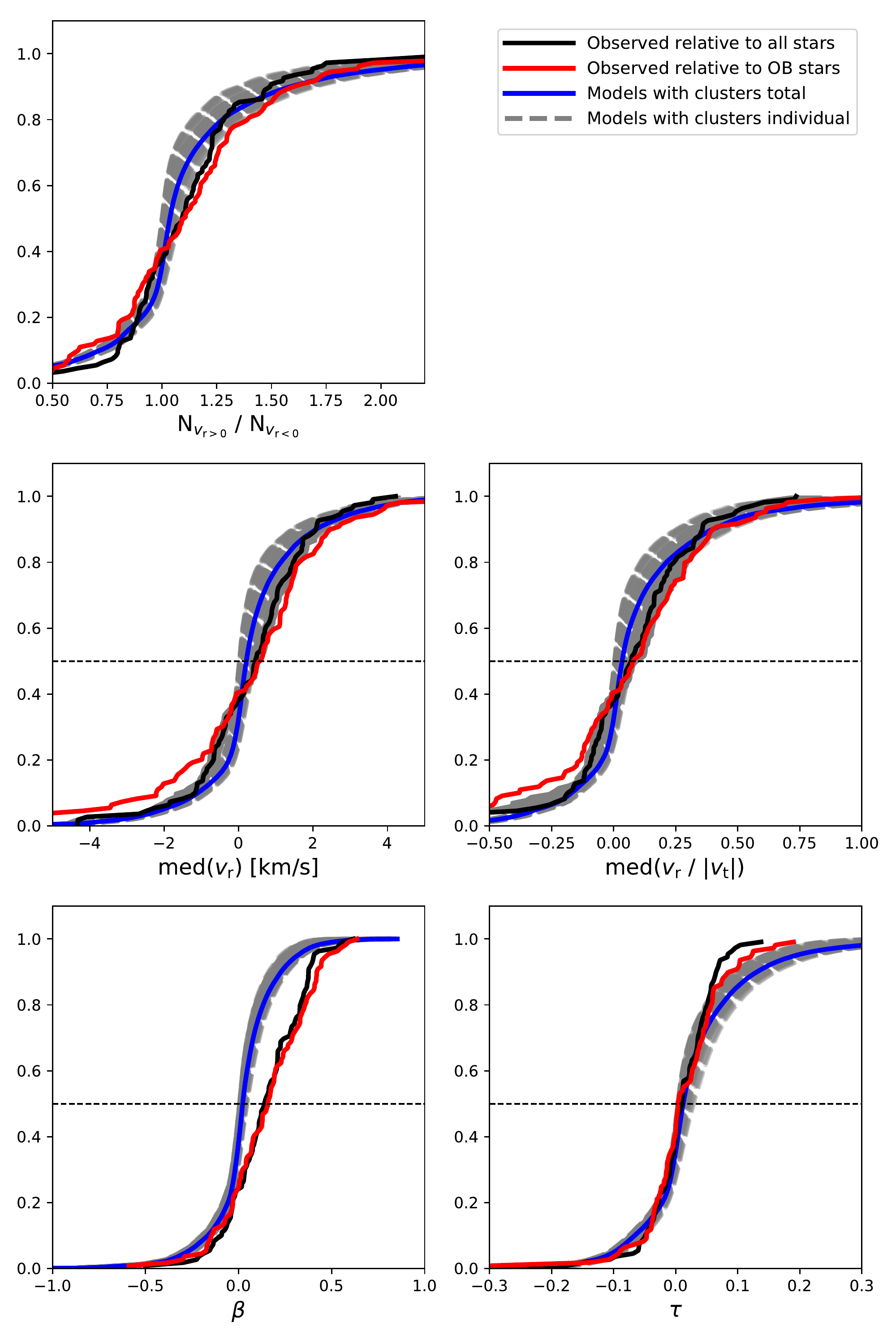}
\end{center}
\caption{\label{6panel_subvel_popclust} Cumulative distributions of each of the five kinematic parameters for model associations with substructure in position and velocity space with expanding clusters included. The distributions for each set of expanding cluster parameters as described in Section \ref{popclusters} are shown in the grey dashed lines with the total combined distribution over all of the models with velocity substructure is shown in the blue solid curve. The cumulative distributions of the observed sample of OB associations are shown in the black and red solid lines. Uncertainties in the observed distribution have been omitted for clarity but can be found in Fig. \ref{DR2_mainplot}. The ratio N$_{v_{\text{r}}>0}/$N$_{v_{\text{r}}<0}$ is shown in the upper-left panel, the median radial velocity in the middle-left panel and the median value of radial velocity normalised by tangential velocity in the middle-right panel. The lower-left panel shows the cumulative distributions for the radial anisotropy parameter ($\beta$) and the lower-right panel shows the distributions for Kendall's $\tau$ measured for radial velocity against radius using the centre of all stars in each association. The observed distributions for almost all kinematic diagnostics are well reproduced by models with positional and velocity substructure with the addition of small-scale localised expansion events, with the exception of the radial-anisotropy parameter which is slightly elevated compared with the model associations. Additionally, the models overestimate the degree to which radial-velocity and radius are correlated in the most positive region of the parameter space. Nonetheless, this set of models provides a satisfactory match to the observed kinematics, considering the uncertainties.
}
\end{minipage}
\end{figure*}

 The resulting cumulative distributions for each of the five measured kinematic properties are shown in Fig. \ref{6panel_subvel_popclust}, alongside the observed cumulative distributions for the same quantities (shown as solid black and red curves). The total cumulative distributions over all models included in the grid are shown in blue. 
 
These models, which include both positional substructure and velocity substructure as well as a small number of radially-anisotropic expansion events, are the first models to remain consistent with observations over the entire distributions for the first three parameters examined (the number ratios and median radial velocities). The models depart from the observed distributions relative to the centre of all stars at the most extreme end of the expanding regime of the associations, however, they remain consistent with the cumulative distributions using the mean position and velocity of OB-type stars as the association centres. The most significant exception to this behaviour is that there is a small ($<0.1$) excess in the observed radial-anisotropy parameter distribution relative to the model distributions. For the cumulative distribution of Kendall's $\tau$ for $v_{\text{r}}$ and radius, the observed distributions exhibit a deficit of radial velocity sorting with respect to the model distributions in the region $\tau > 0$. Even for the kinematic diagnostics that are successfully reproduced by the model distributions, no single combination of the free parameters that make up the model distributions reproduces the complete sample of observed OB associations. Instead, a range of velocity substructure and cluster properties are required.

Given that the expansions seeded in the associations are already maximally radially-anisotropic, simply increasing the degree of radial-anisotropy of expansions is not possible without also increasing the median radial velocity parameters. It is therefore likely that this excess in radial anisotropy originates from compressive motion, that takes place alongside the localised expansion events. One of the most likely sources of this additional component is the hierarchical merging of individually expanding clusters. Increased radial-anisotropy may also result from tidal disruption (e.g. \citealt{Kruijssen2011}), or on the larger-scale, galactic shear affecting the kinematics of the original molecular cloud (e.g. \citealt{Dobbs2012,Jeffreson2018}).

\section{Discussion}

\label{discussion}

We have extended the original study of the kinematics of OB associations in \citet{Paper1} from 18 nearby OB associations to over 109 OB associations using the {\it Gaia}-DR2 astrometric catalogue in conjunction with the GALOBSTARS catalogue. We determine key kinematic properties of these associations and quantify the degree to which they can be considered to be undergoing expansion in excess of what can be expected from a randomised velocity distribution, and what properties must be considered in order to reproduce observations.

\subsection{Implications of this work}

As in \citet{Paper1}, a simple monolithic cluster that subsequently underwent gas-expulsion-driven expansion is firmly ruled out as an origin of the associations examined in this work. However, the kinematic properties of the sample of OB associations analysed here are also found to be inconsistent with randomised velocity fields, unlike the 18 associations of \citet{Paper1}. The main source of this apparent discrepancy is the sample size for each association. The associations appeared to exhibit random velocity fields from the DR1 data, because a random velocity field and a highly substructured velocity field are largely indistinguishable when low numbers of stars are examined. However, with the larger sample supplied by {\it Gaia}-DR2, randomised velocity fields tend towards median radial velocities and radial-anisotropy parameter values of zero, while substructured velocity fields behave similarly to a smaller sample size due to stars being comoving with their neighbours. Additionally, the improved precision and sample size of {\it Gaia-}DR2 has allowed the detection of low levels of expansion within OB associations, likely the result of the expansion of individual substructures within associations.

The majority (2/3) of OB associations exhibit positive median radial velocities with respect to the association centres, defined as the mean position and velocity of all stars within each association. $\sim$80\% of OB associations exhibit elevated levels of radial anisotropy above $\beta = 0$. However, none of these associations are consistent with the levels of anisotropy expected from $N$-body simulations of expanding clusters \citep{Baumgardt2007}.

The lack of a clear correlation between radial velocity and radius for any of the associations presented in this work suggests that while radial velocities are indeed positive on average for most associations, this expansion has not yet had a significant impact on the large-scale structure of the associations. Rather, the expansion of clusters must impact the small-scale substructure of associations while the size of the association is likely to be set by the scale, structure, and dynamical state of the star forming gas at the onset of star formation. This is consistent with a hierarchical picture of star formation in which stars are formed across a wide variety of environments across a star forming region, rather than a limited number of gravitationally-bound clusters \citep[e.g.][]{Kruijssen2012b,Krumholz2019}.

Only with a combination of small-scale localised expansion events, positional substructure and a randomised but substructured velocity field are we able to reproduce the kinematic properties of observed OB associations examined in this work. As demonstrated in Section \ref{improved_models}, each of these elements is necessary for reproducing the variety of kinematic diagnostics considered here across the observed population of OB associations. This therefore rules out simplistic models in which OB associations are formed from a single or multiple expansion of previously compact clusters. It also rules out the purely random velocity field models, that were able to describe the {\it Gaia-}DR1/TGAS observations. Instead, the {\it Gaia}-DR2 data present a much more nuanced picture of OB associations as large scale, highly substructured groups of stars in which some, but by no means all, stars form in compact clusters and groups that undergo small-scale expansion. While the expansion of compact substructures is indeed significant enough to be measured, the large-scale structure of associations is set by other processes and is most likely inherited from the structure and kinematics of the molecular cloud from which the stars originally formed. The excess in radial anisotropy detected is indicative that OB associations exhibit collapsing and expanding behaviour simultaneously. This is possibly the result of hierarchical merging of substructures within the associations.

The investigation of \citet{Melnik2017} of the kinematics of 18 nearby OB associations find that present-day OB associations are gravitationally unbound but find only low levels of expansion. This is entirely consistent with \citet{Paper1} in which we found that the low levels of expansion of OB associations are consistent with randomised velocity fields with the presence of positional substructure. We show in this work that with {\it Gaia}-DR2, OB associations typically exhibit low levels of expansion and radial-anisotropy in excess of what would be expected from randomised velocity fields. 

The recent work of \citet{Kuhn2019} focussed on young stellar objects with ages 1--5\,Myr in 28 clusters and associations using {\it Gaia}-DR2. They find that the majority (75\%) of structures are apparently expanding with a median expansion velocity of 0.4\,km\,s$^{-1}$. This is entirely consistent with the results of this study, matching the median value of our distribution of median radial velocities (0.4\,km\,s$^{-1}$, see Section \ref{results_section}). \citet{Kuhn2019} also find that in NGC 6530, expansion velocity is correlated with radius; however, there is significant scatter in this relation and it is the only stellar group that exhibits a statistically significant relation between radial velocity and radius. Crucially, in this work  we demonstrate that the expansion signatures identified by \citet{Kuhn2019} are entirely consistent with a hierarchical view of star formation in which OB associations are formed in-situ without the requirement for gas-expulsion-driven expansion, and that the expansions on the scale of association substructures are detectable in the bulk properties of the associations (see Fig. \ref{6panel_subvel_popclust}) but do not affect the large-scale structure of the associations (see Fig. \ref{example_Vr_R} and Fig. \ref{6panel_subvel_popclust}).

This, in combination with the lack of a relation between radial velocity and radius for the vast majority of observed OB association across {\it Gaia-}DR2 studies, and the implied insignificance of global expansion, is in agreement with the detailed studies of the OB associations Cygnus OB2 \citep{Wright2014,Wright2015,Wright2016} and Sco Cen \citep{Wright2018} that establish that these associations exhibit significant substructure and have undergone little dynamical evolution. Our results are also consistent with the findings of \citet{Cantat-Gaudin2018}, showing that the expansion of Vela OB2 began before the stars were formed and therefore must be a relic of the initial gas dynamics rather than gas-expulsion-driven expansion. Together, these works create a picture in which it is the fractal structure and complex dynamics of the interstellar medium that drives the evolution of stellar groups rather than monolithic gas-expulsion-driven expansion events.

\subsection{Caveats and limitations}

We use the distance estimates derived by \citet{Bailer-Jones2018}. These are based only on the parallaxes from {\it Gaia}-DR2. As such, it is likely that for many stars in our sample, more precise determinations of distance could be made with the inclusion of additional ancillary data. It is also possible that this approach may include unknown systematic effects, especially as the {\it Gaia} parallaxes may have significant dependencies on colour. Indeed, the fact that a small number of the most distant associations appear to exhibit very high distance dispersions compared to the majority of the sample would suggest that these are likely to be unresolved along the line of sight. However, variations in distance estimates would primarily manifest as increased dispersions in both position and velocity (which we do not rely on for the analysis presented) rather than a bias against finding gas-expulsion-driven expansion-like velocity fields. While changes in distance can alter the absolute velocities derived, it does not change the direction of motion and therefore the radial velocities are the only properties we investigate that may be significantly affected. We also note that the models that we compare the observations to are built using the observed velocity dispersions. This makes the comparison entirely insensitive to distance uncertainties. Moreover, the \citet{Bailer-Jones2018} catalogue of distances offers the largest all-sky catalogue of homogeneously derived distances to date, and as such remains the most suitable source of distances for this study.

Throughout this study, we assume that the high mass OB stars and the lower mass stars within each association share a common centre of mass. Recent work by \citet{Armstrong2018} has shown that this may not be the case in all associations. However, as shown in Fig. \ref{DR2_mainplot}, the choice between defining the centre of an OB association based on the positions and velocities of all stars versus only using the most massive stars in each association has little effect on the measured kinematic properties, except in the most extreme cases.

We make no attempt to measure or assess the gravitational-boundedness of any of the OB associations studied in this work. It is therefore possible that some of the groupings of stars that we have designated associations are in fact gravitationally-bound stellar clusters, and therefore there would be no expectation of systematic expansion in these sources. However, given the range of sizes and morphologies of associations visited with this work (see Table \ref{app_tab1}), it is unlikely that gravitationally-bound structures form a significant proportion of our sample. The only clear contaminant is association 98. This is only 2-3\,pc in size, with a velocity dispersion of $\sim$1\,km\,s$^{-1}$, and both its position in the plane of the sky and distance are consistent with the open cluster NGC 2516.

Importantly the results of this study are limited to the origin of gravitationally-unbound associations. Many stars do form in gravitationally-bound clusters but we do not quantify the cluster formation efficiency in this work \citep[see e.g.][]{Kruijssen2012b,Adamo2015,Johnson2016}.
Moreover, it is worth pointing out that associations can and will dissolve into the Galactic field over time; however, this does not resemble what would be expected from gas-expulsion-driven expansion. Rather, it seems likely that the dissolution of associations is driven by the combination of ballistic dispersal and galactic shear.

 \section{Conclusions}
 
 \label{conclusions}
 
 In this work, we investigate the kinematic properties of a large number of independently selected OB associations. Using the OPTICS clustering algorithm, we select 109 likely OB associations from a large sample of OB stars within the Milky Way.
 
 In contrast to \citet{Paper1}, we find that the majority of observed associations exhibit velocity fields that are inconsistent with random velocity fields. Rather than randomised velocity fields, we conclude the velocity fields are highly substructured, an effect that was not readily distinguishable from randomised velocity fields in the TGAS data due to an insufficient number of stars. This limitation is conclusively overcome by the greatly enhanced statistics in {\it Gaia}-DR2. In addition, a small expanding component is necessary to reproduce the observed kinematic properties of OB associations. However, the lack of any strong correlation between radial velocity and radius indicates that these expansion do not set the overall size or the large-scale structure of the associations.

 While many stars form in gravitationally-bound clusters, and some clusters likely do undergo a well ordered expansion, our results show that OB associations do not form through the expansion of clusters. Localised expansion of individual substructures within associations does appear to be an important component of the kinematic properties of associations. However, this expansion is not the primary driver of their large-scale structural evolution. Therefore, our results are far more consistent with a scale-free, hierarchical picture of star formation, in which stars are formed across a continuous density distribution throughout molecular clouds, rather than exclusively within clusters, and in which OB associations are formed in-situ as relatively large-scale and gravitationally-unbound structures.
 
 Our results falsify the picture that most stars form in clusters, of which a large fraction is subsequently unbound by gas-expulsion-driven expansion. Historically, this idea was largely motivated by $N$-body simulations treating gas expulsion with a modified gravitational potential term. Recent simulations with a self-consistent treatment of hydrodynamics have shown that the star formation efficiency peaks within dense clusters, rendering gas expulsion-driven expansion ineffective. This requires OB associations to have formed in-situ, with a high degree of substructure expected to trace the position-velocity structure of the gas from which the stars formed. Our analysis of the Gaia-DR2 data unambiguously favours this interpretation \citep[e.g.][]{Kruijssen2012b,Krumholz2019}.
 
 Finally, we conclude that in order to reproduce the bulk kinematic properties of OB associations, the presence of substructure (both in position and velocity space) must be taken into account. Even with these features taken into account, there remains a small, and unaccounted for excess in the radial-anisotropy of the velocity field. This could be the result of hierarchical merging of substructures within associations, or external forces such as tidal encounters.

The origin and dynamical evolution of associations is a complex problem with no single one-size-fits-all solution. The reason for this is clear: the gas from which stars are formed is itself a complex dynamical system \citep[e.g.][]{henshaw19}. The molecular clouds from which stars, clusters, and associations are formed are supersonically turbulent, as well as subject to tidal forces, galactic shear, spiral arm interactions, cloud-cloud collisions, and external stellar feedback (e.g. supernovae) before and during the onset of star formation, leading to a wide range of kinematic properties. These are irreproducible by a simple monolithic model of star formation. Future work linking Gaia data to sub-mm observations of nearby star-forming regions will enable a better understanding the strong link between the interstellar medium and young stellar associations.

\section*{Acknowledgements}

We thank the anonymous referee for their useful comments and suggestions.
We also thank Marina Kounkel and Nicholas Wright for their helpful comments and stimulating discussions.
JLW, JMDK, and HWR acknowledge support from the Deutsche Forschungsgemeinschaft (DFG, German Research Foundation) -- Project-ID 138713538 -- SFB 881 (\textquotedblleft The Milky Way System\textquotedblright, subproject B2).
JMDK gratefully acknowledges funding from the DFG in the form of an Emmy Noether Research Group (grant number KR4801/1-1), from the European Research Council (ERC) under the European Union's Horizon 2020 research and innovation programme via the ERC Starting Grant MUSTANG (grant agreement number 714907). This work has made use of data from the European Space Agency (ESA)
mission {\it Gaia} (\url{https://www.cosmos.esa.int/gaia}), processed by
the {\it Gaia} Data Processing and Analysis Consortium (DPAC,
\url{https://www.cosmos.esa.int/web/gaia/dpac/consortium}). Funding
for the DPAC has been provided by national institutions, in particular
the institutions participating in the {\it Gaia} Multilateral Agreement.
The authors acknowledge support by the High Performance and Cloud Computing Group at the Zentrum f\"{u}r Datenverarbeitung of the University of T\"{u}bingen, the state of Baden-W\"{u}rttemberg through bwHPC
and the DFG through grant no INST 37/935-1 FUGG.




\bibliographystyle{mnras}
\bibliography{GAIA_bibliography} 




\appendix

\section{Tables of association properties}

In this appendix we include tables for all associations included in our study detailing the positions and sizes of the observed associations as well as the derived kinematic properties. In Table A1, we list the number of stars included in our sample for each association (which is by no means complete), as well as their positions in the plane in the sky. We also list the most likely corresponding literature association based on the catalogue of \citet{Melnik2009}. We show the mean distance of all selected members for each association, as well as the standard deviations in distance, physical position in the tangential plane at the listed coordinates, and velocities in that plane.
Table A2 lists the derived kinematic properties for each of the observed OB associations, both relative to the mean position and velocity of all stars and relative to the mean position and velocity of only the OB-type stars in each association.

	\begin{table*}
		\caption{\label{app_tab1} Table of association positions and sizes. We list here the OB associations identified in Section 2 by their arbitrarily assigned association numbers. In column 2, we list the OB association from the catalogue of \citet{Melnik2009} that represents the most likely association that meets the criteria that the association falls within 1$\sigma$ (as measured in this study) of the literature position in the plane of the sky and  within 3$\sigma$ of the literature distance. Note that these association names are only given as a guide for the reader, and they do not influence this work in any way. The number of stars and OB stars in each association are listed in the third and fourth column, respectively. This is followed by the positions in the plane of the sky in columns 5 and 6. Column 7 lists the average distance to the selected members of each association. The standard deviations in distance, physical position in the tangential plane at the listed coordinates, and velocities in that plane are shown in the last five columns}
\begin{minipage}{170mm}
		\begin{tabular}{l l c c c c c c c c c c}
		\hline
			Assoc. no. & Assoc. Name & N$_{\ast}$ & N$_{OB}$ & RA & Dec & Dist. & $\sigma_{\text{Dist.}}$  &  $\sigma_{X}$ & $\sigma_{Y}$ & $\sigma_{v_{X}}$ & $\sigma_{v_{Y}}$ \\
			           &        & &   & [$\deg$] & [$\deg$] & [pc] & [pc] & [pc] & [pc] & [km\,s$^{-1}$] & [km\,s$^{-1}$]  \\
			\hline
1	&		&	622	&	8	&	54.33	&	11.08	&	120	&	9	&	16	&	10	&	9	&	11	\\
2	&		&	312	&	9	&	20.00	&	58.59	&	3145	&	75	&	38	&	23	&	8	&	4	\\
3	&		&	1054	&	12	&	8.58	&	62.79	&	3487	&	63	&	164	&	41	&	11	&	7	\\
4	&		&	13895	&	44	&	357.20	&	62.29	&	3601	&	167	&	143	&	76	&	17	&	11	\\
5	&		&	30565	&	25	&	332.40	&	55.02	&	1618	&	123	&	74	&	34	&	15	&	9	\\
6	&		&	2412	&	7	&	334.86	&	55.43	&	3777	&	60	&	54	&	50	&	20	&	16	\\
7	&		&	21855	&	8	&	334.33	&	54.98	&	2298	&	79	&	88	&	35	&	14	&	10	\\
8	&		&	995	&	15	&	335.33	&	55.66	&	4171	&	59	&	110	&	52	&	9	&	7	\\
9	&	Mon OB2	&	63410	&	50	&	100.21	&	5.02	&	1656	&	193	&	54	&	78	&	12	&	14	\\
10	&	Sgr OB5	&	4527	&	16	&	267.36	&	-28.60	&	2415	&	71	&	41	&	42	&	9	&	11	\\
11	&		&	15089	&	11	&	273.58	&	-20.21	&	2253	&	56	&	33	&	56	&	12	&	13	\\
12	&		&	78849	&	10	&	287.38	&	13.90	&	1092	&	80	&	46	&	62	&	14	&	18	\\
13	&		&	16433	&	9	&	241.20	&	-54.40	&	2107	&	50	&	92	&	23	&	16	&	14	\\
14	&		&	42057	&	17	&	234.44	&	-55.14	&	1759	&	86	&	104	&	63	&	16	&	10	\\
15	&		&	4707	&	10	&	259.30	&	-39.56	&	2643	&	65	&	47	&	26	&	16	&	20	\\
16	&		&	31729	&	59	&	297.21	&	23.36	&	1993	&	158	&	56	&	44	&	12	&	11	\\
17	&		&	71463	&	15	&	246.28	&	-51.02	&	1878	&	76	&	96	&	53	&	13	&	13	\\
18	&		&	6660	&	9	&	354.64	&	59.80	&	1702	&	51	&	93	&	48	&	10	&	5	\\
19	&	Gem OB1	&	61206	&	98	&	92.29	&	22.46	&	1832	&	290	&	70	&	57	&	12	&	13	\\
20	&		&	2481	&	10	&	93.05	&	22.80	&	2369	&	43	&	59	&	40	&	9	&	6	\\
21	&		&	2167	&	26	&	116.37	&	-32.29	&	3609	&	124	&	57	&	62	&	7	&	6	\\
22	&		&	21913	&	10	&	245.36	&	-51.59	&	2456	&	55	&	71	&	28	&	19	&	16	\\
23	&	NGC 6204	&	6657	&	11	&	250.35	&	-47.34	&	2350	&	73	&	56	&	36	&	9	&	9	\\
24	&		&	16169	&	15	&	70.14	&	43.55	&	939	&	79	&	73	&	52	&	7	&	6	\\
25	&		&	49004	&	12	&	295.32	&	22.17	&	880	&	65	&	48	&	53	&	11	&	12	\\
26	&	Mon OB1	&	1488	&	8	&	99.80	&	9.39	&	727	&	29	&	14	&	14	&	6	&	4	\\
27	&		&	24684	&	11	&	274.70	&	-14.48	&	934	&	66	&	39	&	24	&	12	&	12	\\
28	&		&	55711	&	15	&	297.24	&	22.49	&	1217	&	84	&	42	&	57	&	13	&	11	\\
29	&		&	4923	&	29	&	264.30	&	-33.69	&	2380	&	179	&	29	&	32	&	9	&	9	\\
30	&		&	26187	&	11	&	33.10	&	60.29	&	989	&	64	&	69	&	52	&	13	&	7	\\
31	&		&	1684	&	21	&	26.05	&	62.06	&	3339	&	61	&	123	&	77	&	11	&	6	\\
32	&		&	3420	&	18	&	7.49	&	62.49	&	3028	&	110	&	88	&	37	&	15	&	7	\\
33	&	Cas OB4	&	3422	&	9	&	5.98	&	62.85	&	2142	&	38	&	71	&	40	&	10	&	4	\\
34	&		&	4662	&	19	&	133.11	&	-46.61	&	2169	&	49	&	89	&	63	&	8	&	7	\\
35	&		&	61207	&	16	&	194.76	&	-60.43	&	1239	&	69	&	83	&	55	&	20	&	8	\\
36	&		&	3876	&	9	&	317.64	&	57.32	&	837	&	57	&	63	&	29	&	6	&	5	\\
37	&		&	30959	&	39	&	106.60	&	-22.86	&	847	&	106	&	50	&	39	&	10	&	10	\\
38	&		&	21688	&	15	&	263.39	&	-34.64	&	410	&	31	&	70	&	44	&	10	&	9	\\
39	&		&	20662	&	11	&	260.80	&	-39.14	&	1946	&	58	&	66	&	61	&	9	&	10	\\
40	&		&	1921	&	9	&	271.48	&	-24.39	&	1408	&	24	&	15	&	32	&	10	&	10	\\
41	&		&	10502	&	11	&	278.02	&	-19.28	&	730	&	59	&	47	&	14	&	8	&	9	\\
42	&		&	105284	&	47	&	178.45	&	-61.58	&	1901	&	140	&	166	&	58	&	19	&	10	\\
43	&		&	208	&	9	&	160.73	&	-60.64	&	3411	&	48	&	43	&	34	&	9	&	4	\\
44	&		&	4262	&	17	&	167.22	&	-60.35	&	3410	&	83	&	70	&	42	&	11	&	8	\\
45	&	Cyg OB9	&	36267	&	20	&	306.19	&	39.04	&	1026	&	54	&	54	&	65	&	15	&	18	\\
46	&		&	13949	&	15	&	180.95	&	-62.17	&	1722	&	67	&	74	&	45	&	12	&	8	\\
47	&		&	9117	&	12	&	133.29	&	-42.00	&	1033	&	80	&	34	&	24	&	10	&	7	\\
48	&	Cyg OB1	&	177657	&	290	&	304.21	&	38.42	&	1643	&	239	&	82	&	108	&	16	&	17	\\
49	&	Cma OB1	&	33824	&	35	&	106.61	&	-10.89	&	1195	&	109	&	47	&	48	&	12	&	12	\\
50	&		&	12937	&	16	&	76.60	&	39.94	&	1117	&	64	&	52	&	50	&	8	&	8	\\
51	&		&	3231	&	13	&	76.93	&	42.72	&	1322	&	44	&	70	&	43	&	7	&	7	\\
52	&		&	196	&	8	&	34.77	&	57.36	&	2368	&	14	&	27	&	14	&	2	&	2	\\
53	&		&	1750	&	11	&	349.18	&	60.00	&	2694	&	78	&	47	&	26	&	15	&	10	\\
54	&		&	3568	&	13	&	348.24	&	59.94	&	3056	&	42	&	101	&	50	&	16	&	10	\\
55	&		&	6761	&	28	&	134.68	&	-47.88	&	1843	&	93	&	61	&	32	&	10	&	9	\\
			\hline
		\end{tabular}
\end{minipage}
	\end{table*}
	\begin{table*}	
\begin{minipage}{170mm}
\textbf{Table A1 (cont.)}

		\begin{tabular}{l l c c c c c c c c c c}
		\hline
			Assoc. no. & Assoc. Name & N$_{\ast}$ & N$_{OB}$ & RA & Dec & Dist. & $\sigma_{\text{Dist.}}$  &  $\sigma_{X}$ & $\sigma_{Y}$ & $\sigma_{v_{X}}$ & $\sigma_{v_{Y}}$ \\
						           &    &    &    & [$\deg$] & [$\deg$] & [pc] & [pc] & [pc] & [pc] & [km\,s$^{-1}$] & [km\,s$^{-1}$]  \\
			\hline
56	&		&	16791	&	15	&	118.86	&	-28.41	&	577	&	56	&	27	&	33	&	9	&	10	\\
57	&	Lac OB1	&	2558	&	13	&	341.40	&	40.77	&	506	&	40	&	29	&	20	&	10	&	4	\\
58	&		&	61632	&	25	&	237.79	&	-54.18	&	627	&	62	&	76	&	48	&	15	&	11	\\
59	&	Sco OB1	&	86539	&	109	&	254.01	&	-40.91	&	1426	&	250	&	58	&	56	&	14	&	12	\\
60	&	Sco OB4	&	29397	&	16	&	261.14	&	-34.78	&	1059	&	49	&	42	&	36	&	15	&	13	\\
61	&		&	910	&	16	&	253.36	&	-41.38	&	1961	&	57	&	30	&	25	&	4	&	5	\\
62	&		&	5079	&	9	&	259.81	&	-36.74	&	1785	&	27	&	51	&	57	&	14	&	13	\\
63	&		&	5850	&	10	&	246.49	&	-48.49	&	1407	&	32	&	49	&	42	&	9	&	9	\\
64	&		&	27567	&	9	&	261.77	&	-41.27	&	588	&	42	&	61	&	20	&	11	&	11	\\
65	&		&	9678	&	11	&	313.00	&	46.20	&	909	&	40	&	54	&	32	&	12	&	11	\\
66	&		&	12981	&	18	&	31.29	&	57.98	&	709	&	50	&	69	&	44	&	13	&	8	\\
67	&	Cam OB1	&	7639	&	10	&	54.28	&	56.47	&	783	&	38	&	64	&	36	&	7	&	10	\\
68	&		&	1019	&	5	&	196.44	&	-61.28	&	1682	&	20	&	62	&	21	&	8	&	5	\\
69	&		&	182	&	11	&	157.17	&	-57.88	&	2946	&	24	&	45	&	19	&	6	&	5	\\
70	&	Per OB1	&	18259	&	124	&	35.38	&	57.58	&	2240	&	184	&	78	&	38	&	11	&	8	\\
71	&		&	1158	&	10	&	43.22	&	58.76	&	2186	&	22	&	54	&	29	&	11	&	6	\\
72	&		&	205	&	7	&	44.57	&	57.42	&	2399	&	22	&	72	&	23	&	5	&	3	\\
73	&		&	454	&	12	&	35.64	&	57.13	&	1847	&	31	&	34	&	11	&	7	&	4	\\
74	&		&	1072	&	13	&	5.63	&	62.30	&	2739	&	46	&	79	&	27	&	12	&	4	\\
75	&		&	5060	&	23	&	356.19	&	61.51	&	2490	&	71	&	97	&	28	&	15	&	8	\\
76	&		&	12992	&	20	&	289.33	&	22.48	&	560	&	32	&	50	&	25	&	9	&	8	\\
77	&		&	16850	&	19	&	341.79	&	61.62	&	424	&	34	&	104	&	33	&	14	&	9	\\
78	&		&	9806	&	11	&	244.46	&	-53.97	&	1029	&	23	&	32	&	56	&	13	&	12	\\
79	&		&	40	&	40	&	253.35	&	-46.16	&	1048	&	45	&	74	&	28	&	6	&	6	\\
80	&		&	4572	&	27	&	260.06	&	-35.48	&	1597	&	108	&	34	&	24	&	11	&	8	\\
81	&		&	3060	&	9	&	50.04	&	61.26	&	1105	&	34	&	60	&	33	&	12	&	10	\\
82	&		&	16987	&	18	&	274.28	&	-18.77	&	1546	&	46	&	31	&	37	&	12	&	12	\\
83	&		&	212	&	13	&	274.50	&	-12.10	&	2164	&	63	&	6	&	11	&	5	&	5	\\
84	&	Sgr OB4	&	3655	&	11	&	273.70	&	-19.00	&	1874	&	39	&	18	&	23	&	9	&	8	\\
85	&		&	4560	&	24	&	166.83	&	-60.29	&	2936	&	62	&	60	&	33	&	13	&	7	\\
86	&		&	353	&	10	&	160.19	&	-60.08	&	3141	&	35	&	50	&	27	&	17	&	9	\\
87	&		&	1136	&	20	&	25.95	&	61.16	&	2538	&	81	&	69	&	20	&	3	&	2	\\
88	&		&	74	&	10	&	38.14	&	61.45	&	2152	&	32	&	12	&	6	&	3	&	3	\\
89	&		&	244	&	13	&	357.53	&	61.95	&	3156	&	28	&	60	&	20	&	11	&	6	\\
91	&		&	25044	&	18	&	298.29	&	39.06	&	234	&	36	&	50	&	36	&	12	&	17	\\
92	&		&	9836	&	14	&	107.48	&	-27.50	&	379	&	24	&	30	&	46	&	9	&	7	\\
93	&	Ori OB1	&	15311	&	85	&	85.26	&	2.59	&	355	&	44	&	20	&	48	&	5	&	8	\\
94	&		&	12737	&	8	&	33.21	&	62.34	&	342	&	18	&	90	&	17	&	24	&	14	\\
95	&		&	23828	&	30	&	300.33	&	36.61	&	372	&	46	&	68	&	26	&	13	&	10	\\
96	&		&	1464	&	7	&	318.29	&	47.19	&	427	&	24	&	23	&	13	&	9	&	9	\\
97	&		&	4308	&	27	&	123.92	&	-47.02	&	387	&	47	&	41	&	22	&	8	&	4	\\
98	& NGC 2516	&	223	&	11	&	119.55	&	-60.68	&	407	&	7	&	3	&	2	&	1	&	1	\\
99	&		&	4015	&	9	&	162.26	&	-58.76	&	485	&	26	&	39	&	13	&	10	&	8	\\
100	&		&	15074	&	19	&	196.11	&	-63.34	&	597	&	36	&	72	&	29	&	23	&	10	\\
101	&		&	19119	&	61	&	190.80	&	-60.50	&	873	&	60	&	115	&	26	&	13	&	8	\\
102	&		&	10005	&	48	&	335.31	&	61.77	&	878	&	54	&	87	&	24	&	8	&	5	\\
103	&		&	6631	&	22	&	325.19	&	57.12	&	973	&	46	&	63	&	29	&	10	&	7	\\
104	&		&	291	&	13	&	167.27	&	-60.42	&	2677	&	23	&	31	&	20	&	6	&	5	\\
105	&		&	1733	&	18	&	62.74	&	19.32	&	149	&	32	&	14	&	15	&	8	&	12	\\
106	&		&	4074	&	7	&	53.08	&	56.98	&	188	&	18	&	31	&	17	&	17	&	9	\\
107	&		&	818	&	8	&	270.63	&	-23.47	&	1118	&	21	&	13	&	21	&	6	&	5	\\
108	&		&	40	&	12	&	274.43	&	-12.00	&	1727	&	24	&	5	&	6	&	3	&	3	\\
109	&		&	40	&	12	&	274.66	&	-13.77	&	1475	&	27	&	4	&	4	&	3	&	3	\\
110	&		&	65	&	11	&	158.75	&	-58.16	&	2515	&	22	&	13	&	5	&	6	&	3	\\

			\hline
		\end{tabular}
\end{minipage}
	\end{table*}

	\begin{table*}
		\caption{\label{app_tab2} Table listing the derived kinematic properties of each OB association in our sample of 109. From left to right: ratio between the number of stars moving outwards over the number of stars moving inwards, median radial velocity, median value of the radial velocity normalised by the magnitude of the tangential velocity, radial anisotropy parameter $\beta$. Values are given using the mean position and velocity of all stars as the association centre and the mean position and velocity of OB stars as the association centre, as indicated in the top row.}
\begin{minipage}{170mm}
		\begin{tabular}{l c c c c c c c c}
		\hline
		           & \multicolumn{4}{|c|}{all stars}         &          \multicolumn{4}{|c|}{OB stars}  \\
			Assoc. no. & N$_{v_{\text{r}}}/$N$_{v_{\text{t}}}$ & $v_{\text{r}}$ & $v_{\text{r}}/|v_{\text{t}}|$ & $\beta$ & N$_{v_{\text{r}}}/$N$_{v_{\text{t}}}$ & $v_{\text{r}}$ & $v_{\text{r}}/|v_{\text{t}}|$ & $\beta$ \\
			\hline
1	&	0.80	&	-1.11	$\pm$	8.77	&	-0.17	$\pm$	1.25	&	-0.064	$\pm$	0.026	&	0.30	&	-7.06	$\pm$	9.28	&	-0.74	$\pm$	1.30	&	-0.375	$\pm$	0.132	\\
2	&	1.14	&	0.40	$\pm$	5.44	&	0.05	$\pm$	1.17	&	-0.136	$\pm$	0.019	&	1.11	&	0.41	$\pm$	6.28	&	0.08	$\pm$	1.25	&	-0.174	$\pm$	0.024	\\
3	&	1.29	&	1.50	$\pm$	10.33	&	0.25	$\pm$	2.07	&	0.348	$\pm$	0.022	&	1.31	&	1.70	$\pm$	10.53	&	0.28	$\pm$	1.93	&	0.331	$\pm$	0.020	\\
4	&	1.46	&	3.60	$\pm$	14.63	&	0.36	$\pm$	1.79	&	0.193	$\pm$	0.003	&	1.97	&	7.10	$\pm$	17.48	&	0.68	$\pm$	2.00	&	0.363	$\pm$	0.006	\\
5	&	0.99	&	-0.07	$\pm$	13.12	&	-0.01	$\pm$	1.91	&	0.330	$\pm$	0.004	&	0.97	&	-0.19	$\pm$	13.11	&	-0.02	$\pm$	1.90	&	0.319	$\pm$	0.004	\\
6	&	1.08	&	0.89	$\pm$	19.95	&	0.07	$\pm$	1.84	&	0.224	$\pm$	0.011	&	1.44	&	4.35	$\pm$	22.54	&	0.39	$\pm$	2.04	&	0.290	$\pm$	0.013	\\
7	&	1.02	&	0.15	$\pm$	13.42	&	0.02	$\pm$	1.89	&	0.329	$\pm$	0.004	&	0.92	&	-0.72	$\pm$	13.56	&	-0.07	$\pm$	1.71	&	0.230	$\pm$	0.003	\\
8	&	0.88	&	-0.65	$\pm$	9.44	&	-0.12	$\pm$	2.07	&	0.350	$\pm$	0.019	&	0.87	&	-0.70	$\pm$	9.98	&	-0.12	$\pm$	2.13	&	0.420	$\pm$	0.025	\\
9	&	0.81	&	-1.84	$\pm$	13.66	&	-0.18	$\pm$	1.65	&	0.107	$\pm$	0.001	&	1.17	&	1.51	$\pm$	15.50	&	0.12	$\pm$	1.46	&	0.004	$\pm$	0.001	\\
10	&	0.97	&	-0.20	$\pm$	10.34	&	-0.02	$\pm$	1.50	&	-0.020	$\pm$	0.001	&	1.11	&	0.64	$\pm$	10.31	&	0.07	$\pm$	1.52	&	0.015	$\pm$	0.001	\\
11	&	0.86	&	-1.14	$\pm$	12.47	&	-0.12	$\pm$	1.47	&	-0.090	$\pm$	0.002	&	0.55	&	-5.31	$\pm$	14.29	&	-0.52	$\pm$	1.63	&	0.198	$\pm$	0.003	\\
12	&	0.96	&	-0.41	$\pm$	15.40	&	-0.03	$\pm$	1.50	&	0.033	$\pm$	0.001	&	1.08	&	0.79	$\pm$	16.37	&	0.07	$\pm$	1.56	&	0.077	$\pm$	0.001	\\
13	&	1.16	&	1.39	$\pm$	15.36	&	0.13	$\pm$	1.57	&	0.112	$\pm$	0.002	&	0.80	&	-2.21	$\pm$	16.80	&	-0.19	$\pm$	1.60	&	0.129	$\pm$	0.002	\\
14	&	1.23	&	1.73	$\pm$	13.73	&	0.19	$\pm$	1.74	&	0.199	$\pm$	0.002	&	1.28	&	2.11	$\pm$	14.00	&	0.22	$\pm$	1.63	&	0.169	$\pm$	0.002	\\
15	&	1.46	&	3.61	$\pm$	17.16	&	0.32	$\pm$	1.76	&	0.210	$\pm$	0.007	&	1.89	&	7.23	$\pm$	17.89	&	0.70	$\pm$	1.95	&	0.445	$\pm$	0.014	\\
16	&	0.94	&	-0.47	$\pm$	10.86	&	-0.05	$\pm$	1.35	&	-0.091	$\pm$	0.001	&	0.80	&	-1.53	$\pm$	11.49	&	-0.16	$\pm$	1.36	&	-0.068	$\pm$	0.001	\\
17	&	1.20	&	1.54	$\pm$	13.63	&	0.16	$\pm$	1.75	&	0.169	$\pm$	0.001	&	0.93	&	-0.68	$\pm$	14.94	&	-0.07	$\pm$	1.74	&	0.176	$\pm$	0.001	\\
18	&	1.10	&	0.59	$\pm$	8.83	&	0.10	$\pm$	1.83	&	0.281	$\pm$	0.008	&	1.22	&	1.14	$\pm$	8.91	&	0.18	$\pm$	1.79	&	0.239	$\pm$	0.006	\\
19	&	0.94	&	-0.43	$\pm$	12.56	&	-0.05	$\pm$	1.50	&	0.075	$\pm$	0.001	&	0.94	&	-0.41	$\pm$	12.38	&	-0.05	$\pm$	1.50	&	0.070	$\pm$	0.001	\\
20	&	0.89	&	-0.54	$\pm$	7.97	&	-0.11	$\pm$	1.71	&	0.210	$\pm$	0.008	&	0.98	&	-0.13	$\pm$	8.21	&	-0.02	$\pm$	1.66	&	0.207	$\pm$	0.008	\\
21	&	1.02	&	0.10	$\pm$	7.00	&	0.02	$\pm$	1.51	&	0.044	$\pm$	0.002	&	0.91	&	-0.26	$\pm$	6.72	&	-0.06	$\pm$	1.49	&	0.013	$\pm$	0.001	\\
22	&	1.23	&	2.14	$\pm$	16.63	&	0.16	$\pm$	1.48	&	-0.028	$\pm$	0.001	&	1.26	&	2.38	$\pm$	16.57	&	0.18	$\pm$	1.48	&	-0.076	$\pm$	0.001	\\
23	&	0.97	&	-0.22	$\pm$	8.82	&	-0.03	$\pm$	1.67	&	0.211	$\pm$	0.005	&	0.97	&	-0.20	$\pm$	9.84	&	-0.03	$\pm$	1.66	&	0.196	$\pm$	0.004	\\
24	&	1.14	&	0.66	$\pm$	6.89	&	0.12	$\pm$	1.65	&	0.122	$\pm$	0.002	&	1.13	&	0.61	$\pm$	7.11	&	0.11	$\pm$	1.71	&	0.185	$\pm$	0.002	\\
25	&	1.15	&	1.05	$\pm$	11.85	&	0.14	$\pm$	1.85	&	0.300	$\pm$	0.003	&	1.24	&	1.53	$\pm$	11.68	&	0.20	$\pm$	1.76	&	0.244	$\pm$	0.003	\\
26	&	0.77	&	-0.65	$\pm$	4.97	&	-0.19	$\pm$	1.35	&	-0.032	$\pm$	0.002	&	0.62	&	-1.68	$\pm$	5.82	&	-0.49	$\pm$	2.10	&	0.099	$\pm$	0.006	\\
27	&	1.11	&	0.74	$\pm$	12.46	&	0.08	$\pm$	1.50	&	0.058	$\pm$	0.001	&	1.17	&	1.23	$\pm$	12.70	&	0.13	$\pm$	1.54	&	0.069	$\pm$	0.001	\\
28	&	1.05	&	0.34	$\pm$	12.72	&	0.04	$\pm$	1.57	&	0.151	$\pm$	0.002	&	1.06	&	0.52	$\pm$	14.10	&	0.05	$\pm$	1.47	&	0.129	$\pm$	0.001	\\
29	&	1.30	&	1.26	$\pm$	8.49	&	0.21	$\pm$	1.47	&	-0.034	$\pm$	0.001	&	1.31	&	1.26	$\pm$	8.63	&	0.23	$\pm$	1.49	&	-0.031	$\pm$	0.001	\\
30	&	0.94	&	-0.35	$\pm$	10.87	&	-0.05	$\pm$	1.63	&	0.141	$\pm$	0.002	&	0.99	&	-0.05	$\pm$	11.65	&	-0.01	$\pm$	1.67	&	0.162	$\pm$	0.002	\\
31	&	1.73	&	2.78	$\pm$	8.50	&	0.49	$\pm$	1.82	&	0.162	$\pm$	0.009	&	1.72	&	2.77	$\pm$	8.52	&	0.50	$\pm$	1.94	&	0.173	$\pm$	0.009	\\
32	&	1.24	&	1.73	$\pm$	12.68	&	0.26	$\pm$	2.21	&	0.375	$\pm$	0.013	&	1.77	&	3.92	$\pm$	12.47	&	0.61	$\pm$	2.14	&	0.419	$\pm$	0.015	\\
33	&	1.26	&	1.34	$\pm$	8.75	&	0.23	$\pm$	2.00	&	0.285	$\pm$	0.009	&	1.30	&	1.61	$\pm$	8.93	&	0.28	$\pm$	2.08	&	0.337	$\pm$	0.011	\\
34	&	1.12	&	0.64	$\pm$	8.21	&	0.12	$\pm$	1.85	&	0.329	$\pm$	0.009	&	1.26	&	1.11	$\pm$	8.18	&	0.21	$\pm$	1.85	&	0.308	$\pm$	0.008	\\
35	&	1.06	&	0.50	$\pm$	15.09	&	0.05	$\pm$	1.66	&	0.122	$\pm$	0.001	&	1.03	&	0.35	$\pm$	15.72	&	0.03	$\pm$	1.64	&	0.158	$\pm$	0.001	\\
36	&	1.22	&	0.71	$\pm$	5.26	&	0.16	$\pm$	1.62	&	0.075	$\pm$	0.002	&	1.30	&	1.12	$\pm$	5.34	&	0.23	$\pm$	1.74	&	0.047	$\pm$	0.001	\\
37	&	0.80	&	-1.28	$\pm$	9.31	&	-0.16	$\pm$	1.30	&	-0.155	$\pm$	0.002	&	0.70	&	-1.64	$\pm$	9.47	&	-0.27	$\pm$	1.40	&	-0.031	$\pm$	0.001	\\
38	&	1.56	&	2.52	$\pm$	9.59	&	0.36	$\pm$	1.61	&	0.018	$\pm$	0.001	&	1.64	&	2.98	$\pm$	9.65	&	0.40	$\pm$	1.56	&	0.040	$\pm$	0.001	\\
39	&	1.31	&	1.61	$\pm$	9.51	&	0.23	$\pm$	1.58	&	0.090	$\pm$	0.001	&	1.45	&	2.23	$\pm$	9.72	&	0.31	$\pm$	1.63	&	0.151	$\pm$	0.002	\\
40	&	0.70	&	-2.25	$\pm$	10.18	&	-0.31	$\pm$	1.69	&	0.061	$\pm$	0.003	&	0.58	&	-3.42	$\pm$	9.03	&	-0.45	$\pm$	1.57	&	-0.046	$\pm$	0.002	\\
41	&	0.82	&	-0.85	$\pm$	7.66	&	-0.12	$\pm$	1.27	&	-0.114	$\pm$	0.002	&	0.82	&	-0.72	$\pm$	7.56	&	-0.13	$\pm$	1.30	&	-0.136	$\pm$	0.003	\\
42	&	0.92	&	-0.78	$\pm$	16.23	&	-0.08	$\pm$	1.88	&	0.375	$\pm$	0.003	&	0.87	&	-1.24	$\pm$	16.42	&	-0.13	$\pm$	1.89	&	0.417	$\pm$	0.003	\\
43	&	0.86	&	-0.93	$\pm$	6.86	&	-0.16	$\pm$	1.62	&	0.145	$\pm$	0.020	&	0.58	&	-2.75	$\pm$	7.07	&	-0.38	$\pm$	1.48	&	-0.168	$\pm$	0.021	\\
44	&	1.07	&	0.40	$\pm$	10.01	&	0.05	$\pm$	1.58	&	0.099	$\pm$	0.003	&	1.19	&	1.30	$\pm$	10.48	&	0.14	$\pm$	1.45	&	0.000	$\pm$	0.001	\\
45	&	0.91	&	-0.95	$\pm$	16.32	&	-0.08	$\pm$	1.67	&	0.206	$\pm$	0.002	&	0.87	&	-1.36	$\pm$	16.21	&	-0.12	$\pm$	1.65	&	0.171	$\pm$	0.002	\\
46	&	1.18	&	1.03	$\pm$	10.07	&	0.15	$\pm$	1.67	&	0.205	$\pm$	0.004	&	1.16	&	0.94	$\pm$	10.46	&	0.13	$\pm$	1.68	&	0.191	$\pm$	0.003	\\
47	&	0.90	&	-0.62	$\pm$	8.83	&	-0.10	$\pm$	1.71	&	0.226	$\pm$	0.005	&	0.90	&	-0.62	$\pm$	8.72	&	-0.10	$\pm$	1.73	&	0.239	$\pm$	0.005	\\
48	&	1.06	&	0.65	$\pm$	17.87	&	0.06	$\pm$	1.88	&	0.341	$\pm$	0.002	&	0.99	&	-0.10	$\pm$	18.25	&	-0.01	$\pm$	1.90	&	0.352	$\pm$	0.002	\\
49	&	0.79	&	-1.87	$\pm$	12.21	&	-0.20	$\pm$	1.53	&	-0.034	$\pm$	0.001	&	0.91	&	-0.97	$\pm$	15.31	&	-0.09	$\pm$	1.63	&	0.040	$\pm$	0.001	\\
50	&	1.11	&	0.45	$\pm$	7.84	&	0.07	$\pm$	1.44	&	-0.042	$\pm$	0.001	&	1.09	&	0.43	$\pm$	8.26	&	0.06	$\pm$	1.45	&	-0.034	$\pm$	0.001	\\
51	&	0.73	&	-1.51	$\pm$	7.09	&	-0.26	$\pm$	1.53	&	-0.051	$\pm$	0.002	&	0.89	&	-0.59	$\pm$	8.35	&	-0.09	$\pm$	1.60	&	-0.027	$\pm$	0.001	\\
52	&	1.68	&	0.70	$\pm$	2.12	&	0.56	$\pm$	1.72	&	0.030	$\pm$	0.009	&	0.80	&	-0.32	$\pm$	2.92	&	-0.20	$\pm$	1.20	&	-0.146	$\pm$	0.038	\\
53	&	1.02	&	0.16	$\pm$	13.87	&	0.02	$\pm$	1.78	&	0.210	$\pm$	0.013	&	0.98	&	-0.07	$\pm$	13.49	&	-0.01	$\pm$	1.72	&	0.125	$\pm$	0.008	\\
54	&	0.96	&	-0.32	$\pm$	14.54	&	-0.05	$\pm$	2.02	&	0.367	$\pm$	0.014	&	0.80	&	-2.18	$\pm$	17.69	&	-0.29	$\pm$	2.37	&	0.532	$\pm$	0.018	\\
55	&	0.93	&	-0.52	$\pm$	9.84	&	-0.06	$\pm$	1.46	&	-0.010	$\pm$	0.001	&	0.88	&	-0.73	$\pm$	9.67	&	-0.10	$\pm$	1.45	&	-0.007	$\pm$	0.001	\\

			\hline
		\end{tabular}
\end{minipage}
	\end{table*}
	
		\begin{table*}	
	\begin{minipage}{170mm}
\textbf{Table A2 (cont.)}

		\begin{tabular}{l l c c c c c c c c c}
		\hline
		                        & \multicolumn{4}{|c|}{all stars}         &          \multicolumn{4}{|c|}{OB stars}  \\
			Assoc. no. & N$_{v_{\text{r}}}/$N$_{v_{\text{t}}}$ & $v_{\text{r}}$ & $v_{\text{r}}/|v_{\text{t}}|$ & $\beta$ & N$_{v_{\text{r}}}/$N$_{v_{\text{t}}}$ & $v_{\text{r}}$ & $v_{\text{r}}/|v_{\text{t}}|$ & $\beta$ \\
			\hline
56	&	0.87	&	-0.71	$\pm$	10.04	&	-0.09	$\pm$	1.54	&	0.131	$\pm$	0.002	&	0.85	&	-0.98	$\pm$	10.22	&	-0.12	$\pm$	1.57	&	0.157	$\pm$	0.003	\\
57	&	1.69	&	2.72	$\pm$	7.47	&	0.60	$\pm$	2.24	&	0.368	$\pm$	0.017	&	1.70	&	2.01	$\pm$	7.40	&	0.54	$\pm$	2.22	&	0.448	$\pm$	0.021	\\
58	&	1.27	&	2.01	$\pm$	14.13	&	0.24	$\pm$	1.91	&	0.394	$\pm$	0.004	&	1.18	&	1.42	$\pm$	14.69	&	0.17	$\pm$	2.04	&	0.465	$\pm$	0.005	\\
59	&	1.09	&	0.67	$\pm$	12.57	&	0.07	$\pm$	1.46	&	-0.042	$\pm$	0.001	&	1.07	&	0.66	$\pm$	13.73	&	0.06	$\pm$	1.50	&	-0.040	$\pm$	0.001	\\
60	&	0.90	&	-0.96	$\pm$	14.30	&	-0.08	$\pm$	1.51	&	-0.060	$\pm$	0.001	&	0.60	&	-4.19	$\pm$	14.13	&	-0.38	$\pm$	1.45	&	0.065	$\pm$	0.001	\\
61	&	1.50	&	1.39	$\pm$	5.00	&	0.35	$\pm$	1.96	&	0.163	$\pm$	0.009	&	1.52	&	1.35	$\pm$	4.69	&	0.37	$\pm$	1.86	&	0.147	$\pm$	0.009	\\
62	&	0.90	&	-0.86	$\pm$	12.56	&	-0.09	$\pm$	1.40	&	-0.148	$\pm$	0.006	&	1.47	&	3.13	$\pm$	13.77	&	0.28	$\pm$	1.45	&	-0.151	$\pm$	0.006	\\
63	&	1.61	&	2.10	$\pm$	8.79	&	0.38	$\pm$	1.51	&	0.029	$\pm$	0.001	&	1.47	&	2.69	$\pm$	10.10	&	0.32	$\pm$	1.53	&	0.126	$\pm$	0.003	\\
64	&	1.23	&	1.44	$\pm$	10.74	&	0.16	$\pm$	1.50	&	0.047	$\pm$	0.001	&	1.24	&	1.43	$\pm$	10.51	&	0.16	$\pm$	1.47	&	0.012	$\pm$	0.000	\\
65	&	1.05	&	0.39	$\pm$	12.27	&	0.05	$\pm$	1.63	&	0.159	$\pm$	0.003	&	1.18	&	1.10	$\pm$	11.26	&	0.12	$\pm$	1.37	&	-0.132	$\pm$	0.002	\\
66	&	1.16	&	0.96	$\pm$	10.50	&	0.14	$\pm$	1.69	&	0.174	$\pm$	0.003	&	0.87	&	-0.94	$\pm$	11.55	&	-0.15	$\pm$	1.83	&	0.331	$\pm$	0.005	\\
67	&	1.18	&	0.84	$\pm$	8.10	&	0.14	$\pm$	1.40	&	-0.017	$\pm$	0.001	&	1.39	&	1.81	$\pm$	9.22	&	0.28	$\pm$	1.63	&	0.191	$\pm$	0.004	\\
68	&	1.00	&	-0.05	$\pm$	7.39	&	-0.01	$\pm$	1.96	&	0.275	$\pm$	0.015	&	1.51	&	2.18	$\pm$	7.72	&	0.42	$\pm$	2.30	&	0.355	$\pm$	0.018	\\
69	&	1.22	&	0.38	$\pm$	4.59	&	0.14	$\pm$	1.46	&	-0.038	$\pm$	0.005	&	0.80	&	-0.38	$\pm$	4.75	&	-0.11	$\pm$	1.31	&	-0.159	$\pm$	0.023	\\
70	&	1.46	&	2.08	$\pm$	9.75	&	0.32	$\pm$	1.66	&	0.136	$\pm$	0.002	&	1.56	&	2.22	$\pm$	10.12	&	0.38	$\pm$	1.72	&	0.128	$\pm$	0.002	\\
71	&	1.33	&	1.70	$\pm$	9.47	&	0.32	$\pm$	2.05	&	0.376	$\pm$	0.028	&	1.20	&	1.11	$\pm$	9.82	&	0.19	$\pm$	1.96	&	0.287	$\pm$	0.021	\\
72	&	1.47	&	0.89	$\pm$	4.75	&	0.33	$\pm$	1.91	&	0.399	$\pm$	0.062	&	1.73	&	1.53	$\pm$	5.07	&	0.57	$\pm$	2.22	&	0.385	$\pm$	0.062	\\
73	&	1.20	&	0.49	$\pm$	6.28	&	0.17	$\pm$	1.83	&	0.406	$\pm$	0.044	&	1.15	&	0.47	$\pm$	5.49	&	0.14	$\pm$	1.84	&	0.374	$\pm$	0.039	\\
74	&	1.14	&	1.13	$\pm$	10.30	&	0.19	$\pm$	2.71	&	0.527	$\pm$	0.039	&	1.26	&	1.28	$\pm$	11.21	&	0.31	$\pm$	2.69	&	0.591	$\pm$	0.040	\\
75	&	1.11	&	0.91	$\pm$	13.20	&	0.13	$\pm$	2.27	&	0.542	$\pm$	0.017	&	1.08	&	0.66	$\pm$	14.98	&	0.11	$\pm$	2.39	&	0.549	$\pm$	0.016	\\
76	&	1.03	&	0.19	$\pm$	9.09	&	0.03	$\pm$	1.64	&	0.016	$\pm$	0.001	&	1.09	&	0.48	$\pm$	9.23	&	0.07	$\pm$	1.61	&	-0.030	$\pm$	0.001	\\
77	&	1.14	&	1.07	$\pm$	12.08	&	0.11	$\pm$	1.71	&	0.027	$\pm$	0.001	&	1.03	&	0.18	$\pm$	11.64	&	0.02	$\pm$	1.40	&	0.041	$\pm$	0.001	\\
78	&	0.87	&	-1.12	$\pm$	12.23	&	-0.12	$\pm$	1.51	&	0.082	$\pm$	0.002	&	0.76	&	-2.02	$\pm$	12.39	&	-0.21	$\pm$	1.52	&	0.071	$\pm$	0.001	\\
79	&	1.22	&	1.46	$\pm$	3.91	&	0.15	$\pm$	1.21	&	0.191	$\pm$	0.103	&	1.22	&	1.46	$\pm$	3.91	&	0.15	$\pm$	1.21	&	0.191	$\pm$	0.103	\\
80	&	0.93	&	-0.39	$\pm$	10.05	&	-0.06	$\pm$	1.57	&	0.109	$\pm$	0.003	&	0.97	&	-0.20	$\pm$	10.55	&	-0.02	$\pm$	1.63	&	0.115	$\pm$	0.003	\\
81	&	0.89	&	-0.67	$\pm$	10.31	&	-0.07	$\pm$	1.36	&	-0.189	$\pm$	0.006	&	1.05	&	0.28	$\pm$	10.46	&	0.03	$\pm$	1.33	&	-0.301	$\pm$	0.011	\\
82	&	1.11	&	0.81	$\pm$	12.04	&	0.09	$\pm$	1.60	&	0.086	$\pm$	0.003	&	0.86	&	-1.43	$\pm$	15.07	&	-0.13	$\pm$	1.66	&	0.083	$\pm$	0.002	\\
83	&	1.16	&	0.41	$\pm$	5.28	&	0.12	$\pm$	1.61	&	0.241	$\pm$	0.039	&	1.33	&	0.78	$\pm$	4.94	&	0.29	$\pm$	1.58	&	0.264	$\pm$	0.043	\\
84	&	1.00	&	0.03	$\pm$	8.44	&	0.00	$\pm$	1.51	&	0.070	$\pm$	0.002	&	1.12	&	0.68	$\pm$	8.74	&	0.10	$\pm$	1.50	&	0.024	$\pm$	0.001	\\
85	&	1.03	&	0.15	$\pm$	10.39	&	0.02	$\pm$	1.74	&	0.226	$\pm$	0.007	&	1.13	&	0.86	$\pm$	11.37	&	0.13	$\pm$	1.82	&	0.282	$\pm$	0.009	\\
86	&	0.87	&	-0.66	$\pm$	12.42	&	-0.07	$\pm$	1.57	&	0.034	$\pm$	0.003	&	0.56	&	-5.82	$\pm$	16.42	&	-0.48	$\pm$	1.88	&	0.229	$\pm$	0.020	\\
87	&	1.75	&	0.93	$\pm$	2.89	&	0.52	$\pm$	1.84	&	0.310	$\pm$	0.018	&	1.54	&	0.71	$\pm$	3.01	&	0.36	$\pm$	1.87	&	0.344	$\pm$	0.019	\\
88	&	2.22	&	1.03	$\pm$	2.61	&	0.44	$\pm$	1.54	&	0.110	$\pm$	0.030	&	2.70	&	2.05	$\pm$	2.91	&	1.13	$\pm$	2.60	&	0.417	$\pm$	0.098	\\
89	&	1.28	&	1.70	$\pm$	9.62	&	0.29	$\pm$	2.32	&	0.617	$\pm$	0.105	&	2.34	&	3.98	$\pm$	8.96	&	0.84	$\pm$	2.44	&	0.636	$\pm$	0.094	\\
91	&	1.49	&	3.10	$\pm$	12.75	&	0.28	$\pm$	1.39	&	-0.297	$\pm$	0.007	&	1.60	&	3.78	$\pm$	12.91	&	0.35	$\pm$	1.42	&	-0.197	$\pm$	0.005	\\
92	&	0.62	&	-2.48	$\pm$	7.15	&	-0.39	$\pm$	1.25	&	-0.214	$\pm$	0.005	&	0.50	&	-2.07	$\pm$	7.37	&	-0.47	$\pm$	1.23	&	-0.094	$\pm$	0.002	\\
93	&	0.28	&	-4.33	$\pm$	4.93	&	-1.07	$\pm$	2.07	&	0.208	$\pm$	0.016	&	0.41	&	-2.22	$\pm$	7.00	&	-0.78	$\pm$	1.81	&	0.403	$\pm$	0.030	\\
94	&	0.93	&	-1.16	$\pm$	21.84	&	-0.09	$\pm$	2.05	&	0.326	$\pm$	0.009	&	0.89	&	-1.73	$\pm$	21.53	&	-0.15	$\pm$	2.02	&	0.329	$\pm$	0.009	\\
95	&	1.55	&	3.04	$\pm$	12.55	&	0.48	$\pm$	1.90	&	0.371	$\pm$	0.005	&	1.40	&	2.45	$\pm$	12.97	&	0.38	$\pm$	1.93	&	0.408	$\pm$	0.006	\\
96	&	0.93	&	-0.35	$\pm$	8.21	&	-0.06	$\pm$	1.51	&	-0.072	$\pm$	0.004	&	1.03	&	0.20	$\pm$	8.11	&	0.03	$\pm$	1.61	&	-0.073	$\pm$	0.004	\\
97	&	0.43	&	-4.08	$\pm$	6.33	&	-1.20	$\pm$	3.18	&	0.559	$\pm$	0.015	&	0.50	&	-3.45	$\pm$	6.68	&	-0.87	$\pm$	2.86	&	0.574	$\pm$	0.015	\\
98	&	0.80	&	-0.15	$\pm$	0.81	&	-0.23	$\pm$	1.64	&	0.119	$\pm$	0.016	&	0.69	&	-0.25	$\pm$	0.96	&	-0.30	$\pm$	1.39	&	0.004	$\pm$	0.001	\\
99	&	2.00	&	4.22	$\pm$	8.15	&	0.74	$\pm$	1.53	&	0.277	$\pm$	0.010	&	1.90	&	3.62	$\pm$	9.20	&	0.61	$\pm$	1.82	&	0.366	$\pm$	0.013	\\
100	&	1.23	&	2.11	$\pm$	17.51	&	0.22	$\pm$	1.89	&	0.375	$\pm$	0.006	&	1.25	&	2.33	$\pm$	17.98	&	0.25	$\pm$	1.98	&	0.409	$\pm$	0.007	\\
101	&	0.96	&	-0.36	$\pm$	12.04	&	-0.05	$\pm$	1.91	&	0.336	$\pm$	0.004	&	0.94	&	-0.60	$\pm$	12.56	&	-0.08	$\pm$	1.81	&	0.298	$\pm$	0.004	\\
102	&	1.22	&	1.02	$\pm$	7.47	&	0.20	$\pm$	2.09	&	0.386	$\pm$	0.007	&	1.30	&	1.31	$\pm$	7.42	&	0.29	$\pm$	2.08	&	0.425	$\pm$	0.008	\\
103	&	1.09	&	0.61	$\pm$	9.42	&	0.09	$\pm$	1.75	&	0.304	$\pm$	0.008	&	1.25	&	1.27	$\pm$	9.24	&	0.21	$\pm$	1.71	&	0.265	$\pm$	0.007	\\
104	&	1.33	&	1.18	$\pm$	6.22	&	0.36	$\pm$	1.60	&	0.093	$\pm$	0.011	&	1.20	&	1.11	$\pm$	6.73	&	0.25	$\pm$	1.71	&	0.076	$\pm$	0.008	\\
105	&	0.44	&	-4.32	$\pm$	8.13	&	-0.62	$\pm$	1.28	&	-0.225	$\pm$	0.028	&	0.40	&	-5.59	$\pm$	6.81	&	-0.86	$\pm$	1.64	&	-0.291	$\pm$	0.031	\\
106	&	1.27	&	1.89	$\pm$	14.00	&	0.19	$\pm$	1.64	&	0.431	$\pm$	0.121	&	1.36	&	1.49	$\pm$	13.24	&	0.22	$\pm$	1.56	&	-0.018	$\pm$	0.005	\\
107	&	1.23	&	0.87	$\pm$	5.42	&	0.20	$\pm$	1.32	&	-0.134	$\pm$	0.009	&	1.15	&	0.49	$\pm$	5.30	&	0.12	$\pm$	1.39	&	-0.172	$\pm$	0.011	\\
108	&	4.00	&	1.73	$\pm$	1.27	&	0.73	$\pm$	1.46	&	-0.501	$\pm$	0.145	&	3.00	&	1.40	$\pm$	1.68	&	0.58	$\pm$	1.30	&	-0.589	$\pm$	0.182	\\
109	&	1.35	&	0.44	$\pm$	3.62	&	0.35	$\pm$	2.81	&	0.402	$\pm$	0.162	&	1.50	&	0.80	$\pm$	3.11	&	0.34	$\pm$	1.33	&	0.484	$\pm$	0.201	\\
110	&	0.55	&	-2.67	$\pm$	6.63	&	-1.02	$\pm$	2.59	&	0.357	$\pm$	0.124	&	0.63	&	-3.14	$\pm$	7.82	&	-0.61	$\pm$	2.73	&	0.444	$\pm$	0.151	\\

			\hline
		\end{tabular}
\end{minipage}
	\end{table*}

\bsp	
\label{lastpage}
\end{document}